\documentclass[twocolumn]{aastex7}
\usepackage{amsmath,amssymb}
\usepackage{mathtools}
\usepackage{color}
\usepackage{graphicx}
\usepackage{epstopdf, epsfig}
 \usepackage{amssymb}
\usepackage{comment}
\usepackage{amsmath}
\usepackage{subcaption}
\usepackage{xcolor}
 \usepackage{hyperref}

\newcommand{\D}{\partial}

\newcommand{\mybv}[1]{\boldsymbol{#1}}

\newcommand{\uvec}{\mybv{u}}

\submitjournal{ApJ}

\begin{document}

\title{Statistical State Dynamics of Large-Scale Structure Formation in Shallow Water Magnetohydrodynamic Turbulence}

\author[gname='Eojin',sname='Kim']{Eojin Kim}
\affiliation{Department of Earth and Planetary Sciences, Harvard University, Cambridge MA 02139,
USA}
\email[show]{ekim@g.harvard.edu}  

\author[gname=Brian, sname=Farrell]{Brian F. Farrell}
\affiliation{Department of Earth and Planetary Sciences, Harvard University, Cambridge MA 02139,
USA}
\email{farrell@seas.harvard.edu}

\begin{abstract}
Zonal jets (ZJ) are prominent coherent structures that spontaneously emerge from the background turbulent state in both stellar and planetary atmospheres. Although formation  and maintenance of coherent jets from small scale hydrodynamic turbulence is well-documented, the mechanism underlying this phenomenon remains controversial. The dynamics of the Earth's polar jet and that of the quasi-biennial oscillation of the equatorial stratosphere have been analytically explained using the Statistical State Dynamics (SSD) framework applied to mid-latitude $\beta$-plane and stratified turbulence of the equatorial region,respectively \citep{Farrell 2003}. Extension of SSD to the shallow water equations of the equatorial $\beta$-plane provided a corresponding theory for the dynamics of Jovian jets \citep{Farrell 2009}.  However, the influence of Lorentz forces in the dynamics of a substantial subset of coherent structures observed in both planetary and stellar turbulence motivates the further extension of SSD analysis of coherent structure formation to magnetohydrodynamics (MHD) turbulence. In this work, we apply the SSD framework to shallow water MHD turbulence to study coherent structure dynamics in which both Reynolds and Maxwell stresses are involved.  Perturbative and nonlinear equilibria SSD solutions reveal formation and statistical equilibration of zonal jet-toroidal field structure (ZJTFS) with both fixed point and time-dependent oscillation behavior with implications for understanding coherent structure formation in MHD turbulence including steady jets such as the solar super-rotation and time-dependent phenomena such as the 22 year solar cycle.\\

\end{abstract}

\section{Introduction}\label{sec:intro}
The emergence of coherent jets from incoherent fluctuation velocity fields is a fundamental phenomenon in the dynamics of planetary turbulence. Prominent examples include the steady, banded winds of the gaseous giant planets \citep{Ingersoll1990} and time-dependent phenomena such as the quasi-biennial oscillation (QBO) of Earth’s equatorial stratosphere \citep{Reed1962}. The polar jet of the Earth’s midlatitude atmosphere exhibits  more structural variation than the jets of the gaseous planets, but  polar jet dynamics is also fundamentally governed by a similar feedback loop between the fluctuation field and the mean flow \citep{Robinson1991, Robinson1996, LorenzHartmann2001}.  The self-organization of these jets by interaction between the jet and its associated turbulent Reynolds stresses has been extensively studied using Statistical State Dynamics (SSD) theory. SSD provides a comprehensive theoretical framework for understanding how background turbulence interacts with mean flows to form coherent structures, which was first applied to the formation of two-dimensional $\beta$-plane jets \citep{Farrell 2003, Farrell 2007}. This framework was later extended to shallow-water turbulence, providing insight into the formation of jets, including the banded winds and the equatorial jet of Jupiter \citep{Farrell 2009}.  SSD theory explained why, depending on  initial condition, gas giant equatorial jets can equilibrate in either a prograde or retrograde form.  SSD theory also provided insight into the morphology of coherent structures, including differences in the form of prograde and retrograde jets, explaining why prograde jets are sharp and narrow, whereas retrograde jets
are wider and more rounded \citep{Constantinou2014}. The mechanism enforcing these structural difference is identified using SSD theory to be nonlinear interaction between the first and second cumulant.  Recent work on SSD theory for jet formation includes a study of the Saturn north polar hexagon \citep{Farrellsaturn}. While SSD theory incorporating the hydrodynamics of coherent structure interaction with turbulence provides understanding of jet formation and maintenance explaining many of the observations of jets in the atmospheres of Earth, Jupiter, and other planets, this understanding is not comprehensive because the magnetized nature of some planetary and stellar atmospheres and interiors necessitates an extension to magnetohydrodynamics (MHD). For instance, previous studies of two-dimensional $\beta$-plane MHD turbulence have demonstrated that the introduction of a mean magnetic field can lead to the weakening or total suppression of zonal jets \citep{Tobias 2007}. Subsequently, a theoretical explanation for this magnetic suppression phenomenon was established using SSD \citep{Constantinou 2018}.  

While understanding jet dynamics in two dimensional $\beta$-plane MHD turbulence provides useful implications, its pure two dimensional divergent free velocity and magnetic field precludes mechanisms important in jet dynamics. The presence of zonal velocity shear allows mean toroidal magnetic field to grow from tilting mean poloidal magnetic field. This mechanism is precluded in pure two dimensional turbulence in the absence of an externally imposed or parameterized poloidal magnetic field. 
While 2D divergence-free models preclude physical mechanisms that form the  poloidal magnetic field required for the $\omega$ tilt mechanism
to operate in the absence of an externally imposed or parameterized poloidal magnetic field, 
shallow water magnetohydrodynamics (SWMHD) models bridge this gap. While retaining the dynamical simplicity of shallow water dynamics, these models permit the inclusion of fluctuation-fluctuation magnetic induction effects that produce a poloidal magnetic field in the absence of an externally imposed poloidal field.
In this paper, we develop a theory for the formation of these zonal jet-toroidal field structures (ZJTFS) by extending the SSD framework to SWMHD turbulence. This model allows for the study of ZJTFS formation with both velocity and magnetic field components, including maintenance of coupled ZJTFS vacillations.  This approach permits detailed analysis of the physics underlying the torsional and magnetic field oscillations of the solar cycle and it can be extended to study the dynamics of other coherent structures in which coupling between forces arising from velocity and magnetic fields interact in planetary and stellar turbulence.\\

\section{SWMHD Equations and Boundary Conditions}\label{sec:setup}
We consider the two-dimensional SWMHD equations on a $\beta$-plane. Dependent variables are the horizontal velocity $\mathbf{u}$, the horizontal magnetic field $\mathbf{B}$, and the layer height, $h$. Velocity  components are $u_x$ in the $x$ (zonal) direction and $u_y$ in the $y$ (meridional) direction, with $x$ and $y$ unit vectors denoted $\hat{i}$, $\hat{j}$ respectively. The governing equations are:

%
\begin{equation}
\begin{split}
 \left( \frac{\partial }{\partial t} + \uvec \cdot \nabla\right) \uvec = -g\nabla h +\frac{1}{4 \pi \rho_o}(\mybv{B} \cdot \nabla)\mybv{B} \\
 - (f_0+\beta y) \mybv{\hat{k}} \times \uvec+ \nu_{u}\Delta \mybv{u}-r_u \mybv{u}
 \label{ueq}
 \end{split}
\end{equation}
\begin{equation}
 (\frac{\partial h}{\partial t})+ \nabla \cdot (h \uvec)=-r_h(h-1)+ \nu_{h}\Delta h
 \label{heq}
\end{equation}
\begin{equation}
\frac{\partial \mybv{B}}{\partial t}+ (\mybv{u}\cdot \nabla) \mybv{B}=(\mybv{B} \cdot \nabla) \mybv{u} + \eta \Delta \mybv{B}-r_B \mybv{B}
\label{Beq}
\end{equation}
where $\rho_0$ denotes the fluid density, $\nu_u$ the kinematic viscosity, and $\eta = (\mu_0 \sigma)^{-1}$ the magnetic diffusivity, with $\mu_0$ and $\sigma$ represent the magnetic permeability and electrical conductivity, respectively. The gravitational acceleration is $g$ and the Coriolis parameter is $f = f_0 + \beta y$, with $\beta$ denoting the planetary vorticity gradient. To ensure numerical stability, an artificial viscosity $\nu_h$ is applied to the layer height $h$;  $r_u$, $r_h$, and $r_B$ represent the Rayleigh damping coefficients for the respective fields. \\
Periodic boundary conditions are imposed in the zonal direction. At both meridional boundaries, $y=-1$ and $y=1$, the layer height is set to unity and both the velocity field and magnetic field are set to zero.
\begin{equation}
\mybv{u}(y=1)=0,~h(y=1)=1, ~\mybv{B}(y=1)=0
\end{equation}
\begin{equation}
\mybv{u}(y=-1)=0, ~h(y=-1)=1, ~\mybv{B}(y=-1)=0
\end{equation}

\section{S3T SSD Formulation of SWMHD Turbulence}\label{sec:SSDformulation}
We now express the  SWMHD equations in SSD form. SSD requires first partitioning the dynamics into mean and fluctuation form which requires choosing a Reynolds average operator.
 We choose the zonal average and denote by an overbar an averaged variable so that:
 \begin{equation}
 \overline{\mybv{u}}=[\mybv{u}]_x=\overline{u}_x(y,t) \hat{i}+ \overline{u}_y(y,t)\hat{j}
 \end{equation}
 \begin{equation}
 \overline{h}=[h]_x=\overline{h}(y,t)
 \end{equation}
 \begin{equation}
 \overline{\mybv{B}}=[\mybv{B}]_x= \overline{B}_x(y,t) \hat{i}+\overline{B}_y (y,t)\hat{j}
 \end{equation}.\\

State variables, decomposed into zonal-mean and fluctuation components, are:\\

\begin{equation}
\mybv{u}= \overline{\mybv{u}}(y,t)+\mybv{u}'(x,y,t)
\end{equation}
\begin{equation}
h=\overline{h}(y,t)+h'(x,y,t)
\end{equation}
\begin{equation}
\mybv{B}=\overline{\mybv{B}}(y,t)+\mybv{B}'(x,y,t)
\end{equation}\\


Taking the zonal mean of equations \eqref{ueq}, \eqref{heq}, and \eqref{Beq}, results in mean equations:\\

\begin{equation}
\begin{split}
\frac{\partial \overline{u}_x}{\partial t}
= -\overline{u}_y \frac{\partial \overline{u}_x}{\partial y } - \overline{u'_y \frac{\partial u'_x}{\partial y}}+\frac{\partial \overline{B_x} }{\partial y}\overline{B_y}+\overline{\frac{\partial B_x' }{\partial y} B_y'}
\\+ (f_0 + \beta y) \overline{u}_y+ \nu_{u} \Delta \overline{u}_x
- r_u \overline{u}_x
\label{meanuxeq}
\end{split}
\end{equation}
\begin{equation}
\begin{split}
\frac{\partial \overline{u}_y}{\partial t}
= -\overline{u}_y \frac{\partial \overline{u}_y}{\partial y}-\overline{u_x'\frac{\partial u'_y}{\partial x}}-\overline{u'_y \frac{\partial u'_y}{\partial y}}-\frac{1}{Fr^2} \frac{\partial \overline{h}}{\partial y}\\+\frac{\partial \overline{B_y}}{\partial y}\overline{B_y}+\overline{\frac{\partial B_y'}{\partial x}B_x'} +\overline{\frac{\partial B_y' }{\partial y}B_y'}
- (f_0 + \beta y) \overline{u}_x\\+ \nu_{\mybv{u}}\Delta \overline{u}_y
- r_u \overline{u}_y
\label{meanuyeq}
\end{split}
\end{equation}
\begin{equation}
\frac{\partial \overline{h}}{\partial t}=-\frac{\partial \overline{h} \overline{v}}{\partial y}- \frac{\partial \overline{h'v'}}{\partial y}-r_h(\overline{h}-1)+\nu_{h}\Delta \overline{h}
\label{meanheq}
\end{equation}
\begin{equation}
\begin{split}
\frac{\partial \overline{B_x}}{\partial t}=-\overline{u}_y \frac{\partial \overline{B_x}}{\partial y}-\overline{u_x'\frac{\partial B_x'}{\partial x}}-\overline{u'_y \frac{\partial B_x'}{\partial y}}+\overline{B_y}\frac{\partial \overline{u}_x}{\partial y}\\+\overline{B_x'\frac{\partial u_x'}{\partial x}}+ \overline{B_y'\frac{\partial u_x'}{\partial y}}+ \eta \Delta \overline{B}_x-r_B \overline{B}_x
\label{meanBxeq}
\end{split}
\end{equation}
\begin{equation}
\begin{split}
\frac{\partial \overline{B_y}}{\partial t}=-\overline{u}_y \frac{\partial \overline{B_y}}{\partial y}-\overline{u'_x\frac{\partial B_y'}{\partial x}}-\overline{u'_y \frac{\partial B_y'}{\partial y}}+\overline{B_y}\frac{\partial \overline{u}_y}{\partial y}\\+\overline{B_x'\frac{\partial u'_y}{\partial x}}+ \overline{B_y'\frac{\partial u'_y}{\partial y}}+\eta \Delta \overline{B}_y-r_B \overline{B}_y
\label{meanByeq}
\end{split}
\end{equation}\\

The fluctuation equations result from subtracting the mean equations \eqref{meanuxeq}, \eqref{meanuyeq}, \eqref{meanheq}, \eqref{meanBxeq}, \eqref{meanByeq} from the corresponding full equations \eqref{ueq}, \eqref{heq}, \eqref{Beq} component by component.
Fluctuation variables are expressed using zonal wavenumber.

\begin{equation}
u'_x(x,y,t)= u'_{xk_x}(y,t) e^{ik_xx}
\end{equation}
\begin{equation}
u'_y(x,y,t)= u'_{yk_x}(y,t) e^{ik_xx}
\end{equation}
\begin{equation}
h'(x,y,t)= h'_{k_x}(y,t) e^{ik_xx}
\end{equation}
\begin{equation}
B'_x(x,y,t)= B'_{xk_x}(y,t) e^{ik_xx}
\end{equation}
\begin{equation}
B'_y(x,y,t)= B'_{yk_x}(y,t) e^{ik_xx}
\end{equation}\\

Formulation of the specific SSD that we employ in this study, denoted S3T, proceeds by closing the expansion in cumulants of the SWMHD SSD at second order \citep{Farrell2019}.  To this end we first cast the fluctuation equations in the compact form:\\

\begin{equation}
\frac{\partial \mybv{\phi}'_{k_x}}{\partial t}=\mybv A \mybv{\phi}'_{k_x}+  \mybv{\xi}(t)
\end{equation}\\

in which the fluctuation state vector is:\\

 \begin{equation}
\mybv{\phi'}_{k_x}=\begin{bmatrix}u'_{xk_x}\\u'_{yk_x}\\ h'_{k_x}\\ B_{xk_x}'\\ B_{yk_x}'\end{bmatrix}
\end{equation}\\
and we have replaced the fluctuation-fluctuation nonlinear term in the momentum and induction equations by stochastic noise processes $\mybv{\xi}(t)$ defined as:\\

\begin{equation}
\mybv{\xi}(t)=\begin{bmatrix} \epsilon_{\mybv{u'u'}}^{1/2}\mybv{F}_{\mybv{u}'}\mybv{\xi}_{\mybv{u}'}(t) \\ 0 \\ \epsilon_{\mybv{B'B'}}^{1/2}\mybv{F}_{\mybv{B}'} \mybv{ \xi}_{\mybv{B}'}(t)
\end{bmatrix}
\end{equation}\\
in which appear the
fluctuation structure matrices $\mybv{F}_{\mybv{u}'}$, $\mybv{F}_{\mybv{B}'}$ and vectors of independent temporally delta-correlated stochastic processes $\mybv{\xi}_{\mybv{u}'}(t)$, $\mybv{\xi}_{\mybv{B}'}(t)$. We note that these stochastic noise process also serve to parametrize any exogeneous excitation in the model.\\ 

The individual components of the matrix of the dynamics, $\mybv{A}$, is described in appendix \ref{sec:Aindividual}:\\

Following the S3T SSD formulation presented in \citep{Farrell2019}, we derive a deterministic Lyapunov equation for the ensemble-mean fluctuation covariance, $\mybv{C}$, using only the operator $A$ and a white-in-time stochastic forcing with spatial covariance, $\mybv{Q}$:\\

\begin{equation}\frac{\partial 
\mathbf{C}}{\partial t}=\mybv A \mathbf{C} + \mathbf{C} \mybv A^{\dagger}+ \epsilon \mathbf{Q} 
\end{equation}

\begin{equation}\mathbf{C}=<\mybv{\phi'}_{k_x} \mybv{\phi'}_{k_x}^{\dagger} >\end{equation}\\
where $\dagger$ denotes the Hermitian transpose.
And, the stochastic closure $\mybv{Q}$ is:\\

\begin{equation}
\mybv{Q}= \begin{bmatrix} \epsilon_{\mybv{u}'\mybv{u}'}\mybv{Q}_{\mybv{u}'\mybv{u}'} & 0 & 0\\ 0 & 0 & 0 \\0 & 0 &\mybv{\epsilon}_{\mybv{B}'\mybv{B}'}\mybv{Q}_{\mybv{B'B'}}
\end{bmatrix}
\end{equation}\\

in which the fluctuation excitation structure correlation matrices appear:\\

\begin{equation}
\mybv{Q}_{\mybv{u'u'}}=\mybv{F}_{\mybv{u'}}\mybv{F}_{\mybv{u}'}^{\dagger}
\end{equation}
\begin{equation}
\mybv{Q}_{\mybv{B'B'}}=\mybv{F}_{\mybv{B'}}\mybv{F}_{\mybv{B'}}^{\dagger}
\end{equation}\\

Equations for the mean state can be expressed in compact form as:\\

\begin{equation}  \frac{\partial\mybv{\Gamma}}{\partial t}=\mathbf{G}(\mybv{\Gamma})+ \mathbf{L}(\mathbf{C})
\label{compactmeaneq}\end{equation}\\
where $\mybv\Gamma=[\overline{u}_x, \overline{u}_y, \overline{h}, \overline{B_x}, \overline{B_y}]^{T}$. The individual components of $\mybv{G} (\mybv{\Gamma})$ and $\mybv{L}(\mybv{C})$ are described in appendix \ref{sec:Ggammaindividual}. 

An important observation is that the fluctuation-fluctuation terms, $\mybv{L}(\mybv{C})$, appearing in the mean equation \eqref{compactmeaneq} can be obtained directly from the second cumulant covariance, $\mybv{C}$; For example, 
\begin{equation}
\overline{u'_xu'_y}=diag(\mybv{L}_{u'_x}\mybv{C}\mybv{L}_{u'_y}^{\dagger})
\end{equation}
in which $\mybv{L}_{u'_x}$ and $\mybv{L}_{u'_y}^{\dagger}$ are linear operators.\\

Summarizing, the S3T SSD equations we employ in this study written in compact form are:\\
\begin{equation}
\frac{\partial  \mybv \Gamma}{\partial t}=\mybv G( \mybv \Gamma)+\mybv L(\mybv C)
\label{SSDcompactmean}
\end{equation}
\begin{equation}
\frac{\partial \mybv C}{\partial t}=\mybv A\mybv C+\mybv C\mybv A^{\dagger}+\epsilon \mybv Q
\label{SSDcompactC}
\end{equation}

\section{S3T Stability Formulation}\label{sec:S3Tstabilityformulation}
Assuming an equilibrium S3T state, $(\mybv \Gamma_e, \mybv C_e)$, in which the LHS of \eqref{SSDcompactmean} and \eqref{SSDcompactC} vanish, and following the derivation in \citep{Farrell-Ioannou-2012, Farrell2019}, perturbation equations linearized around the SSD equilibrium state, $(\mybv \Gamma_e, \mybv C_e)$, can be obtained. The resulting linearized perturbation equations are:\\

\begin{equation}
\frac{\partial \delta \mybv{\Gamma}}{\partial t}=\sum_i\frac{\partial \mybv{G}}{\partial \Gamma_i}|_{\mybv{\Gamma}_e}\delta \Gamma_i+ \mybv L(\delta \mybv{C})
\label{S3Tstabilitymeaneq}
\end{equation}

\begin{equation}
\frac{\partial \delta \mybv C}{\partial t}=\mybv A(\mybv \Gamma_e)\delta \mybv C+\delta \mybv A \mybv C_e+ \delta \mybv C\mybv A( \mybv \Gamma_e)^{\dagger}+\mybv C_e\delta \mybv A^{\dagger}
\label{S3Tstabilitycoveq}
\end{equation}
where 
$$\delta \mybv A=\mybv A(\mybv \Gamma_e+\delta \mybv \Gamma)-\mybv A(\mybv \Gamma_e)$$
Equations \eqref{S3Tstabilitymeaneq} and \eqref{S3Tstabilitycoveq} comprise the formulation of the linear perturbation S3T dynamics. 

\section{Dynamics of Fixed-Point ZJTFS}\label{sec:fixedpointjetdynamics}
The S3T SSD of SWMHD turbulence contains the dynamics of SW hydrodynamic turbulence. For a range of parameter values, the S3T SSD for SW hydrodynamic turbulence supports nonlinear finite amplitude fixed-point equilibria characterized by a velocity jet component in the mean flow \citep{Farrell 2003, Farrell 2009,Constantinou2014}.
Assuming convergence to a fixed point equilibrium containing a velocity jet component,  $(\mybv{\Gamma}_e,\mybv{C}_e)$,  has been obtained, it has explicit form:\\ 

\begin{equation}
\mybv{\Gamma}_e=[\overline{u}_x,\overline{u}_y,\overline{h},0,0]^{T}
\label{jetmeanequilibria}
\end{equation}\\
\begin{equation}
\mybv{C}_e=\begin{bmatrix} \langle u_x' u_x'^{\dagger}\rangle & \langle u'_x u_y'^{\dagger} \rangle & \langle u'_x h'^{\dagger}\rangle & 0 & 0 \\  \langle u_y' u_x'^{\dagger}\rangle & \langle u'_y u_y'^{\dagger} \rangle & \langle u'_y h'^{\dagger}\rangle & 0 & 0 \\
\langle h' u_x'^{\dagger}\rangle & \langle h' u_y'^{\dagger} \rangle & \langle h' h'^{\dagger}\rangle & 0 & 0 \\
0 & 0 & 0 & 0 & 0 \\
0 & 0 & 0 & 0 & 0
\end{bmatrix}
\label{jetcovequilibria}
\end{equation}\\

\begin{figure}
\centering{

\includegraphics[width=.75\linewidth]{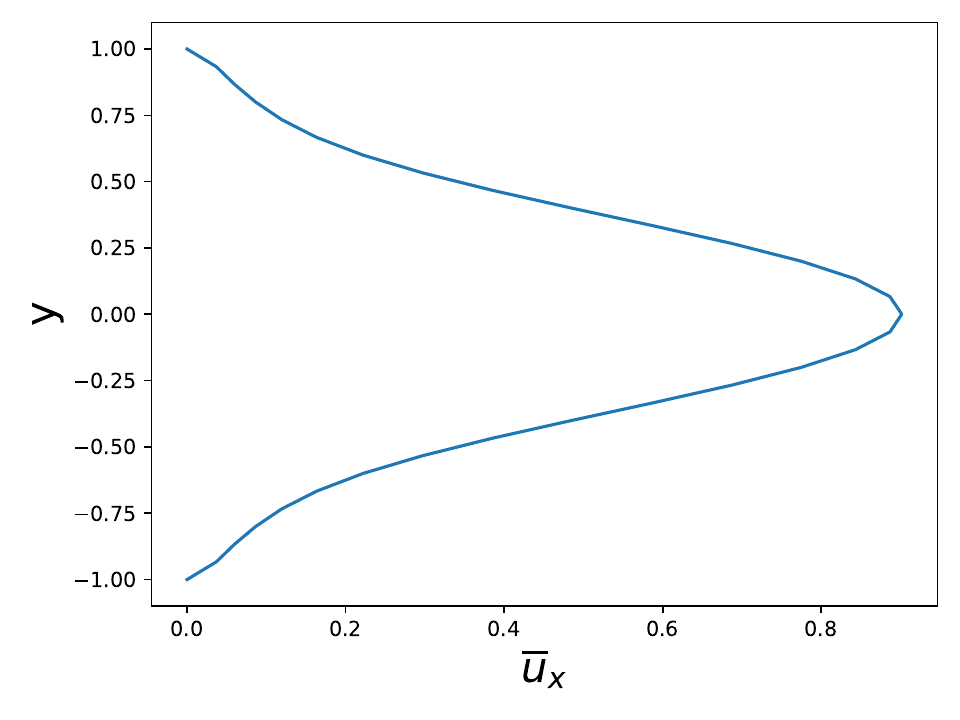}
}
      
\caption{Mean zonal velocity component, $\overline{u}_x$, of the fixed point SW hydrodynamic equilibrium at 
$\epsilon_{\mybv{u}'\mybv{u}'}=0.4, ~\epsilon_{\mybv{B'}\mybv{B}'}=0$,  $\nu_u=0.09,~\nu_h=0.36,~\frac{\nu}{\eta}=0.25$, $r_B=r_u=r_h=0.1$, $f_0=0$, $\beta=1.8$, $g=11.1$.}
\label{fig:zonaljets}
\end{figure}

Stability analysis of this fixed point zonal jet equilibrium proceeds using numerical implementation of the S3T SSD perturbation equations \eqref{S3Tstabilitymeaneq} and \eqref{S3Tstabilitycoveq}. For our analysis we chose a meridional channel size $Ly=2$
 with $Ny=21$ grid points. Dimensionless parameters are $\nu_u=0.09, \nu_h=0.36, $ $r_B=r_u=r_h=0.1$, $f_0=0$, $\beta=1.8$ and $g=11.1$. Values of our dimensionless parameters here are relevant to planets such as Jupiter \citep{Farrell 2009}, whose jets are generally regarded as quasi‑steady rather than strongly vacillating.
Adjacent to the boundaries, we  introduced sponge layers in order to enforce radiation boundary conditions.
Equilibrium structure $(\mybv{\Gamma}_e,\mybv{C}_e)$ and structure of the unstable S3T eigenmodes are influenced by the spatial covariance of the excitation, $\mybv{Q}_{\mybv{u}'\mybv{u}'}$,$\mybv{Q}_{\mybv{B}'\mybv{B}'}$.
We implement the stochastic excitation in stream-functional form where 
\begin{equation}
\mybv{F}_{\mybv{u'}}=[\frac{\partial  F^{\psi}}{\partial y}, -\frac{\partial  F^{\psi}}{\partial x}]^{T}\label{Fueq}\end{equation} where \begin{equation} F^{\psi}(y_i,y_j)=exp(-(y_i-y_j)^2/\delta_f^2)
\label{Fpsieq}
\end{equation}
in order to excite divergent free velocities (c.f. \citep{Farrell 2009}). We choose not to excite the flutuation-flucutation magnetic field directly by taking $\epsilon_{\mybv{B}'\mybv{B}'}=0$ and fix the magnitude of the fluctuation-fluctuation velocity field excitation at $\epsilon_{\mybv{u}'\mybv{u}}=0.4$. We use zonal wavenumber, $k_x$, and magnetic prandtl number, $\frac{\nu}{\eta}$, as stability parameters for our analysis. 
The structure of the mean zonal jet component of the fixed point hydrodynamic equilibria with the zonal wavenumber $kx=6.25$ for the resulting fluctuation-fluctuation covariance is shown in figure \ref{fig:zonaljets}. 

\begin{figure}
\centering{
        \subfloat[]{%
            \includegraphics[width=.75\linewidth]{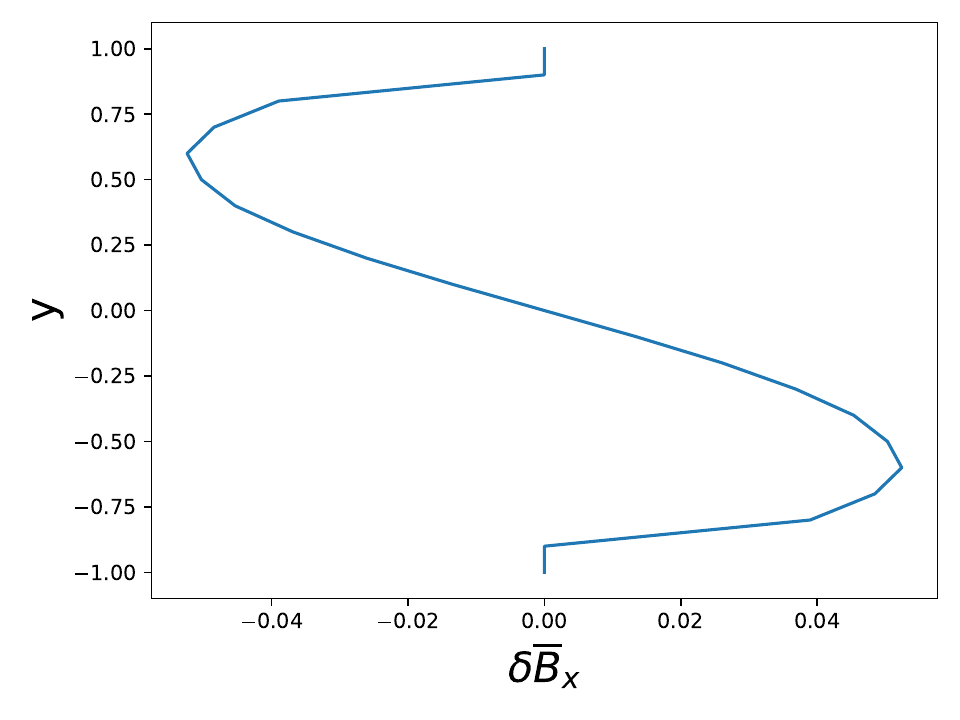}
            \label{subfig:1b}%
        }\\
        \subfloat[]{%
            \includegraphics[width=.75\linewidth]{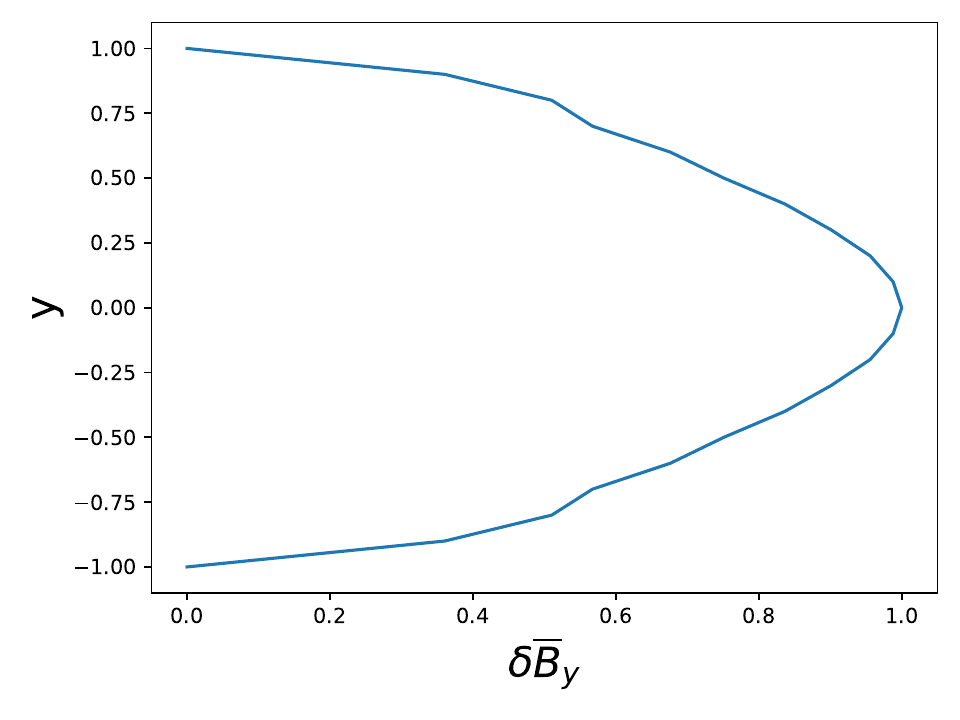}
            \label{subfig:1c}%
        }
        }
      
\caption{Mean magnetic field components of the unstable S3T eigenmode for the equilibrium velocity jet shown in figure \ref{fig:zonaljets}. Panel $(a):$ mean toroidal component, $\delta \overline{B}_x$. Panel $(b):$ mean poloidal component, $\delta \overline{B}_y$. Growth rate $Re(\sigma)=0.76$.  Parameters: $\epsilon_{\mybv{u}'\mybv{u}'}=0.4,~ \epsilon_{\mybv{B'}\mybv{B}'}=0$  $\nu_u=0.09,~\nu_h=0.36,~\frac{\nu}{\eta}=0.25$, $r_B=r_u=r_h=0.1$, $f_0=0$, $\beta=1.8$, $g=11.1$.}
\label{fig:SSDeigenmode}
\end{figure}

In this study, we focus on the ZJTFS dynamics of the prograde jet equilibria. Depending on how the system is initiated, it may alternatively equilibrate to form a retrograde jet.
The structure of the corresponding large scale dynamo instability at Prandtl number $\frac{\nu}{\eta}=0.25$ obtained using \eqref{S3Tstabilitymeaneq} and \eqref{S3Tstabilitycoveq} is shown in figure \ref{fig:SSDeigenmode}. This unstable eigenmode is consistent with the dominant growth mechanism for the mean toroidal field of the eigenmode being tilting of the mean poloidal magnetic field component by the mean shear, $\frac{d \overline{u}_{xe}}{dy}$.
To understand the growth mechanism of the mean perturbation poloidal field component, $\delta \overline{B}_y$,  of the unstable eigenfunction with growth rate Re($\sigma$), the equation for $\delta \overline{B}_y$ is written as:
\begin{equation}
\begin{split}
\frac{\partial \delta \overline{B}_y}{\partial t}=\sigma \cdot \delta \overline{B}_y=-\delta (\overline{u'_x \frac{\partial B_y'}{\partial x}})-\delta (\overline{u'_y \frac{\partial B_y'}{\partial y}})\\+\delta (\overline{ B'_x\frac{\partial u'_y}{\partial x}})+\delta (\overline{B'_y\frac{\partial u'_y}{\partial y}})+\eta \Delta \overline{B}_y-r_B \overline{B}_y
\end{split}
\end{equation}
Forcings of the $\delta \overline{B}_y$ are:
\begin{equation}
\Pi_{\delta\overline{B}_yA}=\eta \Delta \overline{B}_y-r_B \overline{B}_y
\end{equation}
\begin{equation}
\Pi_{\delta \overline{B}_yB}=-\delta (\overline{u'_x \frac{\partial B_y'}{\partial x}})-\delta (\overline{u'_y \frac{\partial B_y'}{\partial y}})
\end{equation}
\begin{equation}
\Pi_{\delta \overline{B}_yC}=\delta (\overline{ B'_x\frac{\partial u'_y}{\partial x}})+\delta (\overline{B'_y\frac{\partial u'_y}{\partial y}})
\end{equation}
These terms are identified as: dissipation, $\Pi_{\delta\overline{B}_yA}$; fluctuation-fluctuation advection, $\Pi_{\delta\overline{B}_yB}$; fluctuation-fluctuation tilting/stretching, $\Pi_{\delta\overline{B}_yC}$.

\begin{figure}

  \centering{

\includegraphics[width=.75\linewidth,height=0.25\textheight]{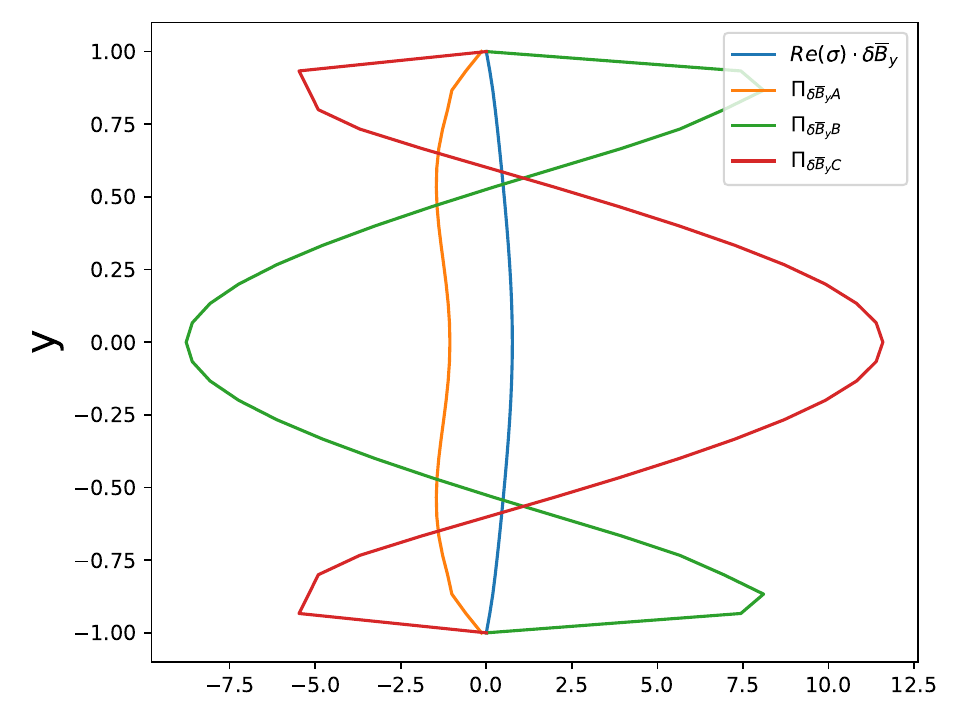}
  
  }
      
\caption{Forcings of the mean poloidal field component, $\delta \overline{B}_y$, of the unstable eigenmode showing growth rate, $Re(\sigma) \cdot \delta \overline{B}_y$, dissipation, $\Pi_{\delta \overline{B}_yA}$, fluctuation-fluctuation advection, $\Pi_{\delta \overline{B}_yB}$, and fluctuation-fluctuation tilting, $\Pi_{\delta \overline{B}_yC}$. $Re(\sigma)=0.76$,
$\frac{\nu}{\eta}=0.25$, $\epsilon_{\mybv{u}'\mybv{u'}}=0.4$, and $\epsilon_{\mybv{B'}\mybv{B}'}=0$.}
\label{fig:LdeltaCcomponent}
\end{figure}

This mean perturbation poloidal field arises from perturbation-perturbation fluxes that are explicitly calculated as shown in figure \ref{fig:LdeltaCcomponent}. Fluctuation-fluctuation stretching/tilting, $\Pi_{\delta \overline{B}_y C}$, provides a positive forcing for $\delta \overline{B}_y$ with an amplitude greater than the combined negative forcings of dissipation, $\Pi_{\delta \overline{B}_y A}$, and fluctuation-fluctuation advection, $\Pi_{\delta \overline{B}_y B}$, in the core region surrounding the jet center. The blue line shows the net growth rate of the torroidal field $\delta \overline{B}_y$, which is $Re(\sigma)\cdot \delta \overline{B}_y$, obtained by summing all three terms. This growth mechanism has in common with the familiar $\alpha$-$\omega$  growth mechanism transfer from toroidal to poloidal magnetic field to sustain field growth, but it differs in providing an explicit solution for a turbulent transfer mechanism while the traditional $\alpha$ effect relies on parameterization of this transfer mechanism.  Specifically, the mean poloidal field component of the unstable eigenmode of equations \eqref{S3Tstabilitymeaneq} and \eqref{S3Tstabilitycoveq} grows against dissipation due to  feedback that is calculated from the fluctuation-fluctuation covariance, $\delta \mybv{C}$. It is remarkable that
despite directly exciting only the fluctuation-flucutation velocity component, $<\mybv{u'}\mybv{u'}^{\dagger}>$, our results show that the fluctuation-fluctuation covariance component, $\delta \mybv{C}$, and the mean flow component, $\delta \mybv {\Gamma}$, cooperatively interact, in the presence of the SW hydrodynamic jet equilbria $(\mybv{\Gamma}_e,\mybv{C}_e)$, to give rise to an unstable mean magnetic field component, $\delta \mybv{\overline{B}}$.

\begin{figure}

        \centering{%
            \includegraphics[width=.75\linewidth]{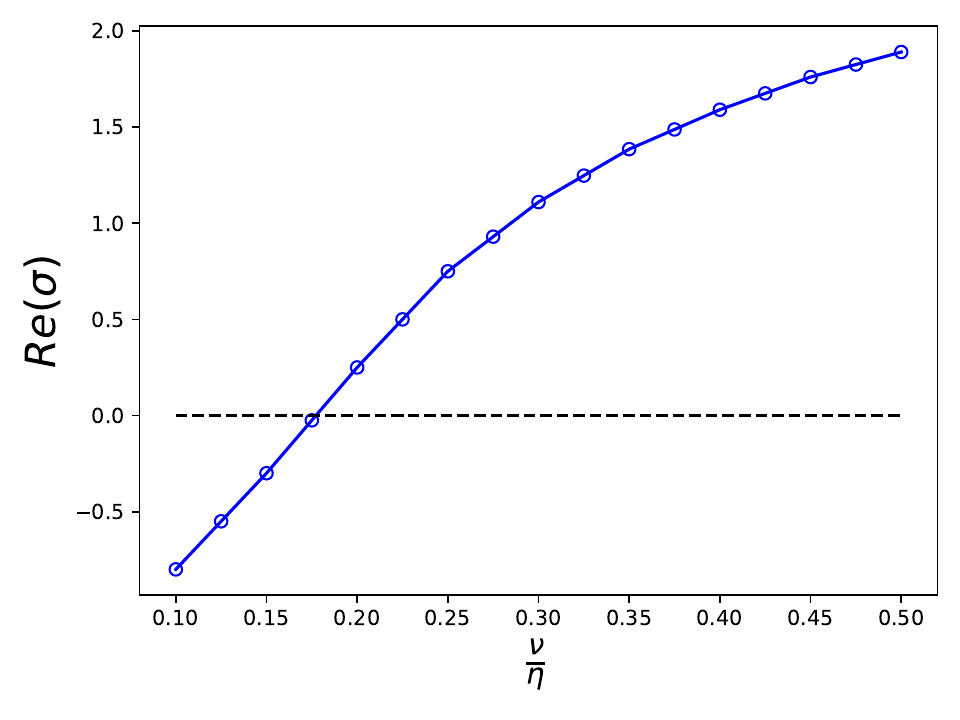}
            
        }
      
\caption{Stability diagram for the large scale dynamo instability of the fixed-point equilibrium velocity jet shown in figure \ref{fig:zonaljets} as a function of magnetic Prandtl number, $\frac{\nu}{\eta}$.  Dashed black line is line of neutral stability. Parameters are: $k_x=6.25$ $\epsilon_{\mybv{u}'\mybv{u}'}=0.4,~\epsilon_{\mybv{B'}\mybv{B}'}=0$, $\nu_u=0.09,~\nu_h=0.36$, $r_B=r_u=r_h=0.1$, $f_0=0$, $\beta=1.8$, $g=11.1$.}
\label{Prandtljetinstability}
\end{figure}

\begin{figure}

        \centering{%
            \includegraphics[width=.75\linewidth]{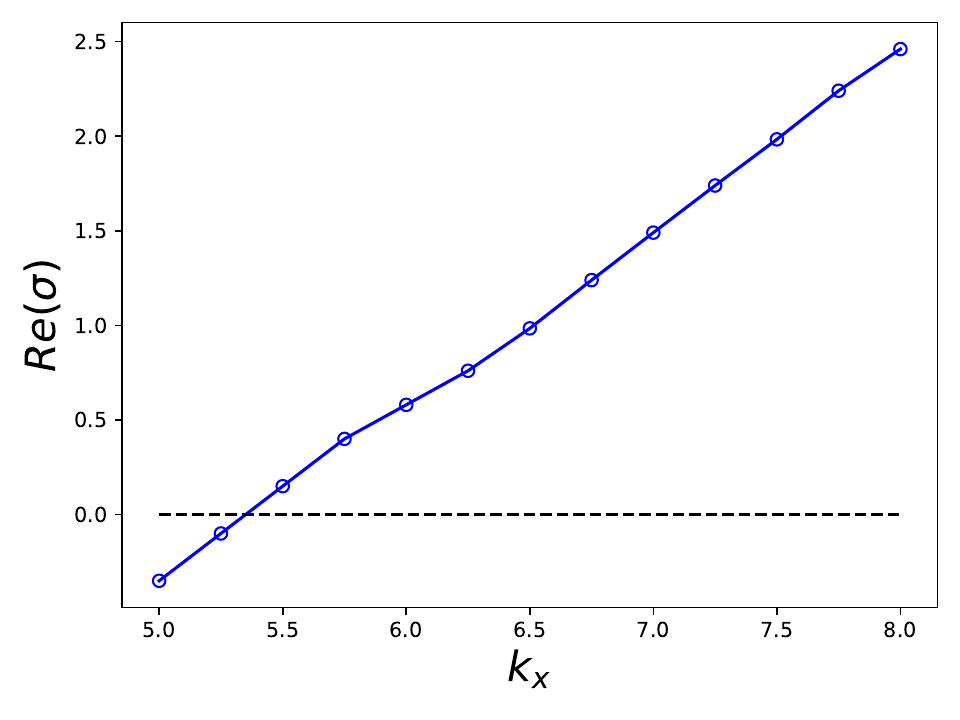}
            
        }
      
\caption{Stability diagram for the large scale dynamo
instability of the fixed-point equilibrium velocity jet shown
in figure 1 as a function of zonal wavenumber, $k_x$.  Dashed black line is the line of neutral stability. Parameters are: $\epsilon_{\mybv{u}'\mybv{u}'}=0.4,~\epsilon_{\mybv{B'}\mybv{B}'}=0$ $\nu_u=0.09,~\nu_h=0.36,~\frac{\nu}{\eta}=0.25$, $r_B=r_u=r_h=0.1$, $f_0=0$, $\beta=1.8$, $g=11.1$.}
\label{kxjetinstability}
\end{figure}

A stability diagram for large scale dynamo formation on the fixed-point jet equilibrium  shown in figure \ref{fig:zonaljets} is shown in figure \ref{Prandtljetinstability} as a function of magnetic Prandtl number, $\frac{\nu}{\eta}$, at zonal wavenumber $kx=6.25$ .
The zonal jet equilibrium  becomes less stable to large scale dynamo instability as $\frac{\nu}{\eta}$ increases,
becoming unstable at approximately  $\frac{\nu}{\eta}\approx 0.175$. 
Another stability parameter to consider is zonal wavenumber $k_x$. 
Shown in figure \ref{kxjetinstability}
is a stability diagram showing the growth rate of the large scale dynamo instability of the zonal jet equilibrium as a function of zonal wavenumber $k_x$ at $\frac{\nu}{\eta}=0.25$. This figure reveals that, given other parameter choices, the zonal jet equilibrium loses stability to large scale dynamo instability at $k_x\approx 5.33$

Stability properties of S3T velocity jet equilibria, defined by \eqref{jetmeanequilibria} and \eqref{jetcovequilibria}, was established using a linear perturbation analysis of the S3T SSD implementation of the SWMHD equations. We now investigate the nonlinear equilibration of these instabilities employing the full S3T SSD system \eqref{SSDcompactmean} -- \eqref{SSDcompactC}.

In order to distinguish equilibrium regimes with and without $\mybv{\overline{B}}$ field, we define the mean toroidal field energy, $E_{\overline{B}_x}$, as:
\begin{equation}
E_{\overline{B}_{x}}=\frac{1}{2}[\overline{B}_x^2]_y
\end{equation}

\begin{figure}

\centering{
        \subfloat[]{%
            \includegraphics[width=.75\linewidth]{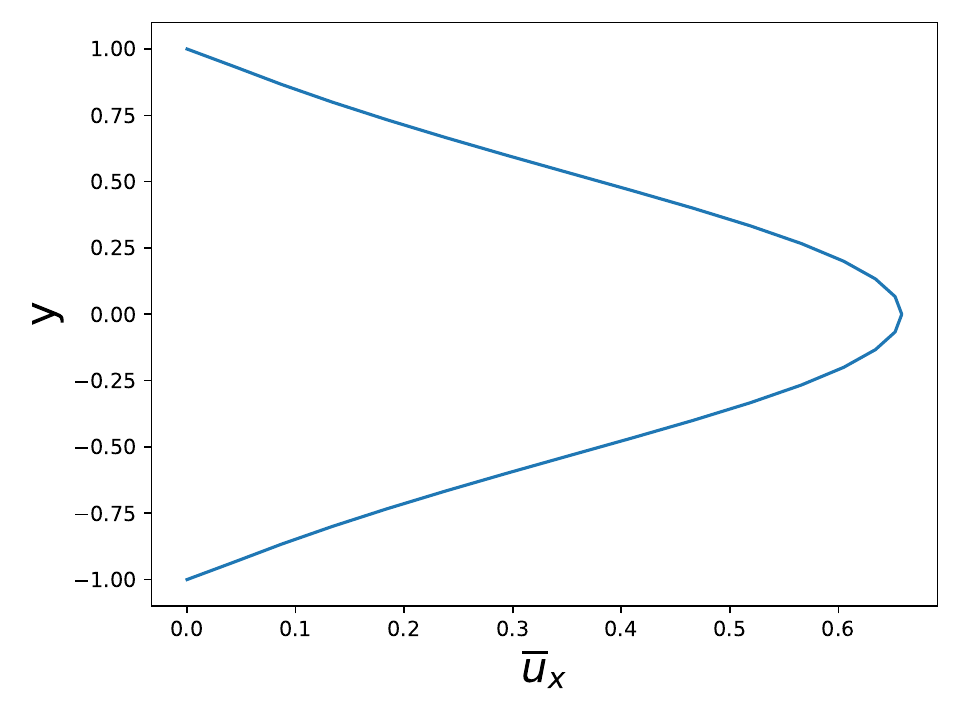}
            \label{subfig:1b}%
        }\\
        \subfloat[]{%
            \includegraphics[width=.75\linewidth]{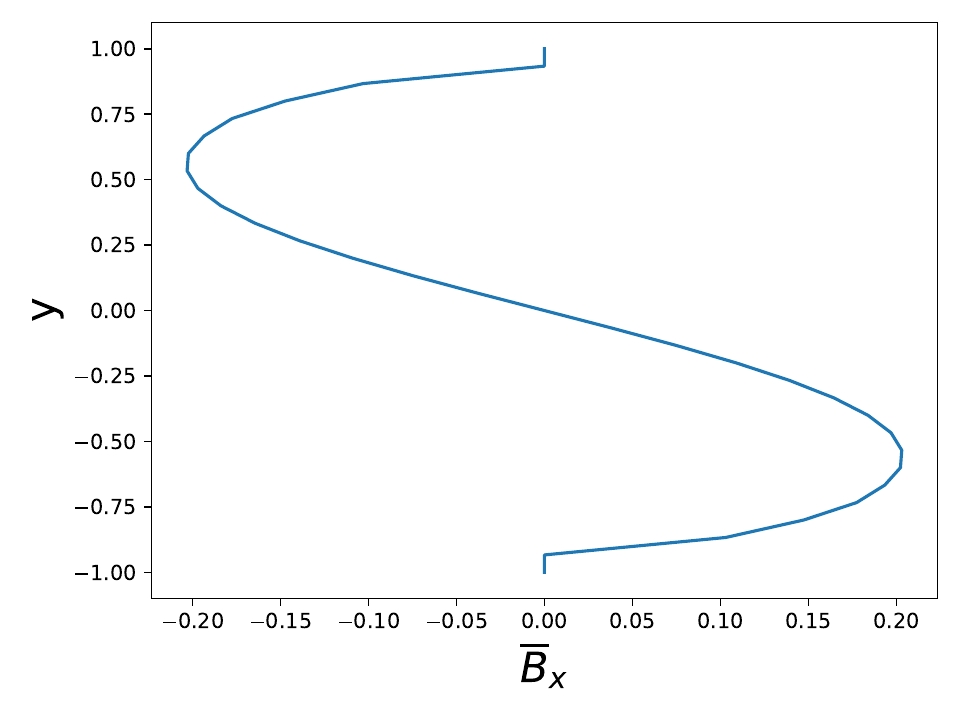}
            \label{subfig:1c}%
        }\\
        \subfloat[]{%
            \includegraphics[width=.75\linewidth]{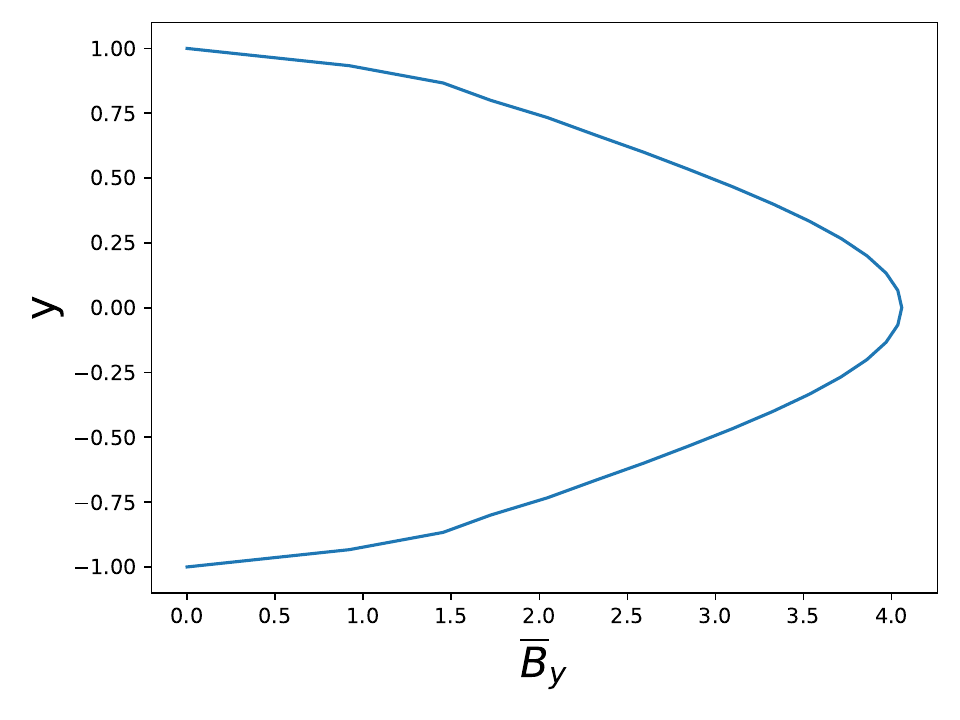}
            \label{subfig:1d}%
        }
        }
      
\caption{Structure of the fixed point equilibrium at $\frac{\nu}{\eta}=0.25$, $\epsilon_{\mybv{u}'\mybv{u}'}=0.4,\epsilon_{\mybv{B}'\mybv{B}'}=0$. Panel $(a)$: zonal jet $\overline{u}_x$ component. Panel $(b)$: mean toroidal magnetic field $\overline{B}_x$. Panel $(c)$:
mean poloidal magnetic field $\overline{B}_y$.}
\label{fig:equilibriumPrmpoint25}
\end{figure}

\begin{figure}

        \centering{
            \includegraphics[width=0.75\linewidth]{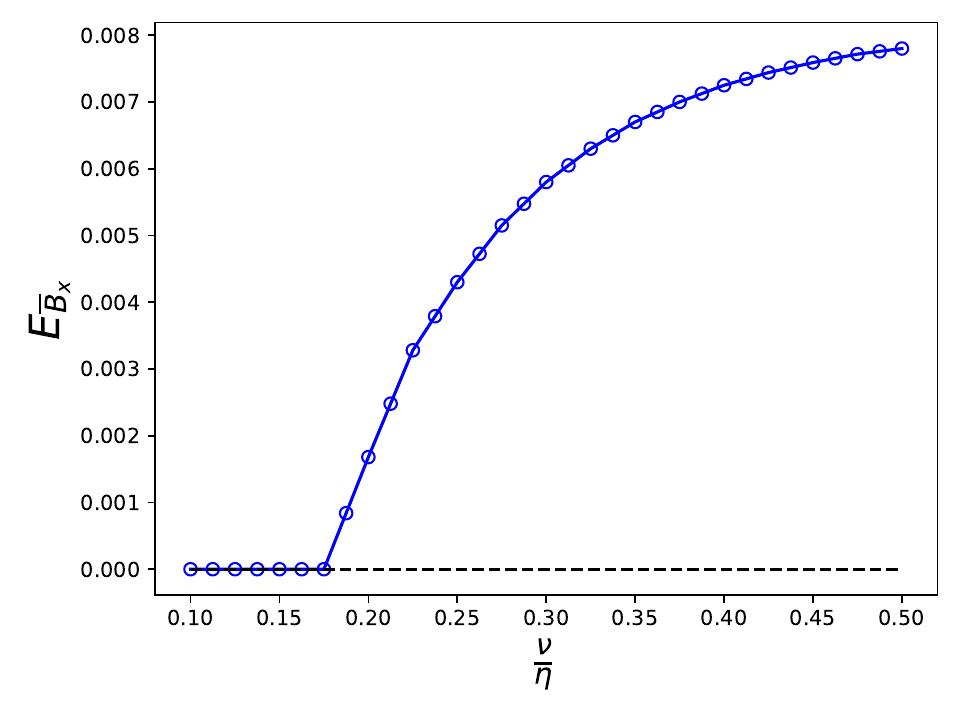}
            
        }
      
\caption{Bifurcation curve showing transition from fixed-point ZJ to fixed-point ZJTFS. Mean toroidal  energy, $E_{\overline{B}_x}$, is shown as a function of Prandtl number $\frac{\nu}{\eta}$. Parameters are: $k_x=6.25$ $\epsilon_{\mybv{u}'\mybv{u}'}=0.4,~\epsilon_{\mybv{B'}\mybv{B}'}=0$ $\nu_u=0.09,~\nu_h=0.36$, $r_B=r_u=r_h=0.1$, $f_0=0$, $\beta=1.8$, $g=11.1$.}
\label{fig:Prmbifurcation}
\end{figure}

This mean toroidal field  is shown  as a function of Prandtl number, $\frac{\nu}{\eta}$, in figure \ref{fig:Prmbifurcation}. 
The transition from a ZJ only fixed-point state to a ZJTFS fixed-point occurs around the Prandtl number $\frac{\nu}{\eta}\approx 0.175$, as shown in figure \ref{fig:Prmbifurcation}.
Shown in figure \ref{fig:equilibriumPrmpoint25} is the structure of a fixed point ZJTFS equilibrium  at $\frac{\nu}{\eta}=0.25$, $\epsilon_{\mybv{u}'\mybv{u}'}=0.4,\epsilon_{\mybv{B}'\mybv{B}'}=0$. This equilibrium state  exhibits dipole structures in the $\overline{B}_y$ component along with antisymmetry about $y=0$ for the $\overline{B}_x$ component. Comparing the maximum amplitude of the zonal jet in the absence of $\overline{\mybv{B}}$ (shown in figure \ref{fig:zonaljets}) with that of the zonal jet in the presence of  $\overline{\mybv{B}}$ (shown in panel $(a)$ of figure \ref{fig:equilibriumPrmpoint25}), the kinetic energy associated with $\overline{u}_x$ has decreased, which is consistent with energy transfer from $\overline{u}_x$ to $\overline{B}_x$ supporting the dynamo.


\section{Mechanisms Maintaining Fixed-Point ZJTFS Equilibria}
When finite amplitude ZJTFS are formed and maintained within the the mean state, $\mybv{\Gamma}$, geostrophic balance is approximately preserved at scales larger than the Rossby deformation radius.
Under these conditions, the equation governing the mean zonal velocity can be expressed as:\\ 

\begin{equation}
\frac{\partial \overline{u}_x}{\partial t}= - \overline{\frac{u'_y \partial u'_x}{\partial y}}+\overline{B}_y\frac{\partial \overline{B}_x}{\partial y}+\overline{B_y'\frac{\partial B_x' }{\partial y}}+\nu_{u}\Delta \overline{u}_x-r_u \overline{u}_x.
\label{ueqformaintenance}
\end{equation}\\

The equation for the associated kinetic energy is obtained by multiplying \eqref{ueqformaintenance}  by $\overline{u}_x$:\\

\begin{equation}
\begin{split}
\frac{\partial \frac{1}{2}(\overline{u}_x)^2}{\partial t}= \overline{u}_x \cdot (- \overline{\frac{u'_y \partial u'_x}{\partial y}})+\overline{u}_x\cdot (\overline{B}_y\frac{\partial \overline{B}_x}{\partial y})\\+\overline{u}_x \cdot (\overline{B_y'\frac{\partial B_x' }{\partial y}})+\overline{u}_x\cdot (\nu_{u}\Delta \overline{u}_x-r_u \overline{u}_x)
\end{split}
\end{equation}\\

Contributions to the mean zonal kinetic energy  are:\\

\begin{equation}
I_{\overline{u}_xA}=(\overline{u}_x)\cdot (\nu \Delta \overline{u}_x-r_u \overline{u}_x) 
\end{equation}
\begin{equation}
I_{\overline{u}_xB}=(\overline{u}_x) \cdot \overline{B}_y \frac{\partial \overline{B}_x}{\partial y}
\end{equation}
\begin{equation}
I_{\overline{u}_xC}=(\overline{u}_x)\cdot -\overline{u'_y \frac{\partial  u'_x}{\partial y}}
\end{equation}
\begin{equation}
I_{\overline{u}_xD}=(\overline{u}_x) \cdot \overline{B'_y\frac{\partial B'_x}{\partial y}.}
\end{equation}\\

 These terms are identified as: dissipation, $(I_{\overline{u}_xA})$; mean magnetic tension contribution, $(I_{\overline{u}_xB})$; Reynolds stress contribution, $(I_{\overline{u}_xC})$; Maxwell stress contribution, $(I_{\overline{u}_xD})$.\\
 
\begin{figure}

       \centering{
            \includegraphics[width=\linewidth,height=0.375\textheight]{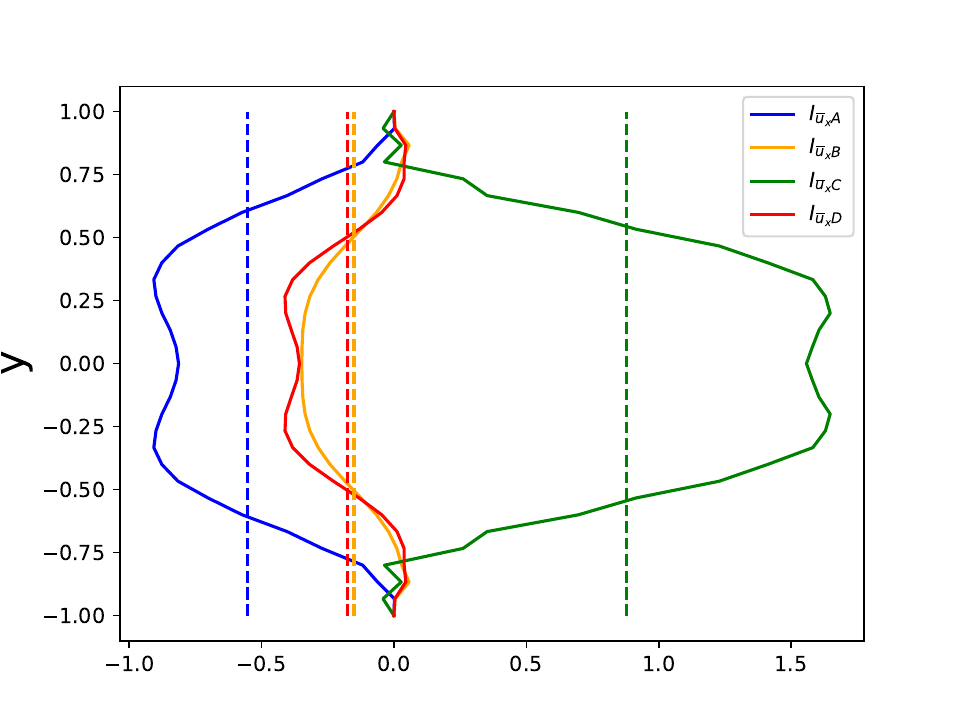}
            
        }

\caption{Equilibrium balance for the mean zonal kinetic energy, $\frac{\overline{u}_x^2}{2}$, showing the dominant Reynolds stress contribution, $(I_{\overline{u}_xC})$, balanced primarily by dissipation, $(I_{\overline{u}_xA})$, with smaller mean magnetic tension force, $(I_{\overline{u}_xB})$,  and  Maxwell stress  contributions, $(I_{\overline{u}_xD})$. Parameters for the fixed-point equilibrium shown are  $\frac{\nu}{\eta}=0.025$, $\epsilon_{\mybv{u}'\mybv{u}'}=0.4,~ \epsilon_{\mybv{B}'\mybv{B}'}=0,~kx=6.25$.}
\label{Ubalancefixedpoint}
\end{figure}

The balance among these terms maintaining the equilibrium mean zonal kinetic energy, 
$\frac{1}{2}(\overline{u}_x)^2$, 
for the equilibrium of the fixed point at 
$\frac{\nu}{\eta}=0.25$,
$\epsilon_{\mybv{u}'\mybv{u}'}=0.4,~\epsilon_{\mybv{B}'\mybv{B}'}=0,~kx=6.25$, 
is shown in figure \ref{Ubalancefixedpoint}.  
The dominant balance is between a positive contributions from Reynolds stress $(I_{\overline{u}_xC})$, opposed by dissipation $(I_{\overline{u}_xA})$, with smaller negative contributions from mean magnetic tension $(I_{\overline{u}_xB})$, and Maxwell stress $(I_{\overline{u}_xD})$.

 We now examine the  mean toroidal field equation:\\

\begin{equation}
\begin{split}
\frac{\partial \overline{B}_x}{\partial t}=-\overline{u'_y \frac{\partial B_x'}{\partial y}}-\overline{u_x'\frac{\partial B_x'}{\partial x}}+\overline{B}_y\frac{\partial \overline{u}_x}{\partial y}+\overline{B_x'\frac{\partial u'_x}{\partial x}}\\+ \overline{B_y'\frac{\partial u'_x}{\partial y}}+ \eta \Delta \overline{B}_x-r_B \overline{B}_x
\label{Bxeqformaintenance}
\end{split}
\end{equation}\\

The equation for the mean toroidal field energy is obtained by multiplying \eqref{Bxeqformaintenance} by $\overline{B}_x$,

\begin{equation}
\begin{split}
\frac{\partial \frac{1}{2}(\overline{B}_x)^2}{\partial t}=\overline{B}_x\cdot (-\overline{u'_y \frac{\partial B_x'}{\partial y}}-\overline{u_x'\frac{\partial B_x'}{\partial x}})\\+\overline{B}_x\cdot (\overline{B}_y\frac{\partial \overline{u}_x}{\partial y})+\overline{B}_x\cdot (\overline{B_x'\frac{\partial u'_x}{\partial x}}+ \overline{B_y'\frac{\partial u'_x}{\partial y}})\\+ \overline{B}_x\cdot (\eta \Delta \overline{B}_x-r_B \overline{B}_x)
\end{split}
\end{equation}\\

Contributions to the mean toroidal field energy are:\\

\begin{equation}
I_{\overline{B}_xA}=(\overline{B}_x)\cdot (\eta \Delta \overline{B}_x -r_B \overline{B}_x)
\end{equation}
\begin{equation}
I_{\overline{B}_xB}=(\overline{B}_x)\cdot (\overline{B}_y\frac{\partial \overline{u}_x}{\partial y})
\end{equation}
\begin{equation}
I_{\overline{B}_xC}=(\overline{B}_x)\cdot (-\overline{u'_x\frac{\partial B'_x}{\partial x}}-\overline{u'_y\frac{\partial B'_x}{\partial y}})
\end{equation}
\begin{equation}
I_{\overline{B}_{x}D}=(\overline{B}_x)\cdot (\overline{B'_x\frac{\partial u'_x}{\partial x}}+\overline{B'_y\frac{\partial u'_x}{\partial y}})
\end{equation}\\ 

These terms are identified as: dissipation, $(I_{\overline{B}_xA})$; mean tilting/stretching, $(I_{\overline{B}_xB})$; fluctuation-fluctuation advection, $(I_{\overline{B}_xC})$; fluctuation-fluctuation tilting/stretching, $(I_{\overline{B}_xD})$. 

\begin{figure}
\centering{
\includegraphics[width=\linewidth,height=0.375\textheight]{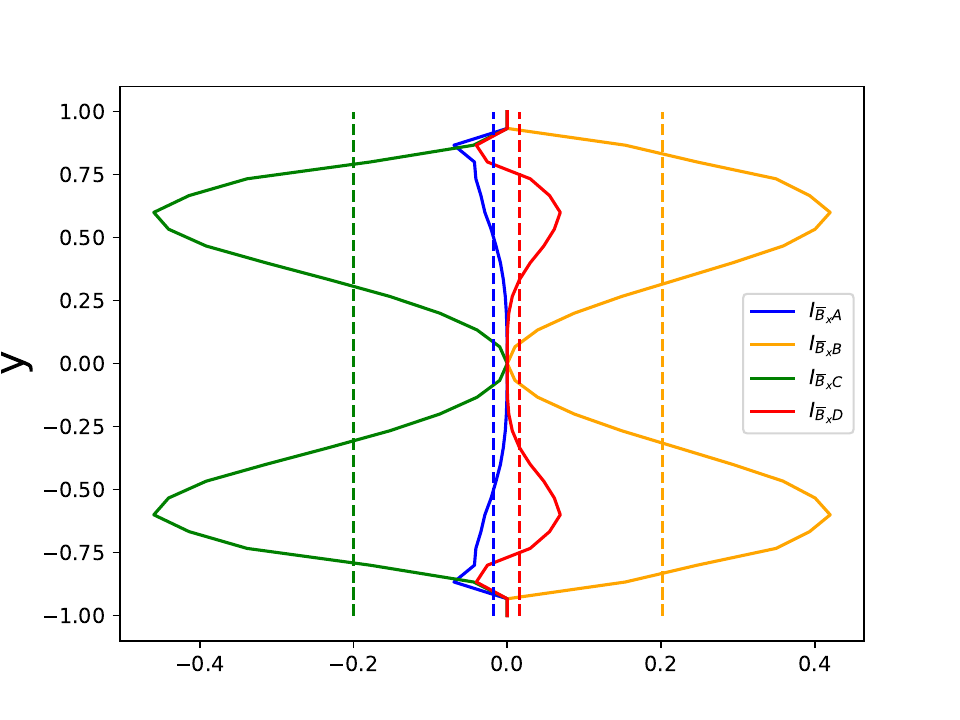}
}
\caption{
Equilibrium balance for the mean toroidal field energy, $\frac{\overline{B}_x^2}{2}$ , showing the dominant balance consists of a positive contribution from mean tilting/stretching, $(I_{\overline{B}_xB})$, counteracted by a negative contribution from fluctuation-fluctuation advection, $(I_{\overline{B}_x C})$, with dissipation, $(I_{\overline{B}_xA})$,   and the fluctuation-fluctuation tilting/stretching, $(I_{\overline{B}_xD})$, making minor contributions. Parameters for the fixed-point equilibrium shown are  $\frac{\nu}{\eta}=0.025$, $\epsilon_{\mybv{u}'\mybv{u}'}=0.4,~ \epsilon_{\mybv{B}'\mybv{B}'}=0,~kx=6.25$.
}
\label{Bxbalancefixedpoint}
\end{figure}

The quilibrium balance maintaining the mean toroidal field energy, $\frac{1}{2}(\overline{B}_x)^2$, for a fixed point equilibrium at $\frac{\nu}{\eta}=0.25$, $\epsilon_{\mybv{u}'\mybv{u}'}=0.4,~ \epsilon_{\mybv{B}'\mybv{B}'}=0,kx=6.25$ is shown in figure \ref{Bxbalancefixedpoint}. The dominant balance consists of a positive contribution from mean tilting/stretching, $(I_{\overline{B}_xB})$, opposed by a negative contribution from fluctuation-fluctuation advection, $(I_{\overline{B}_x C})$.

 \begin{figure}
\centering{
\includegraphics[width=\linewidth,height=0.375\textheight]{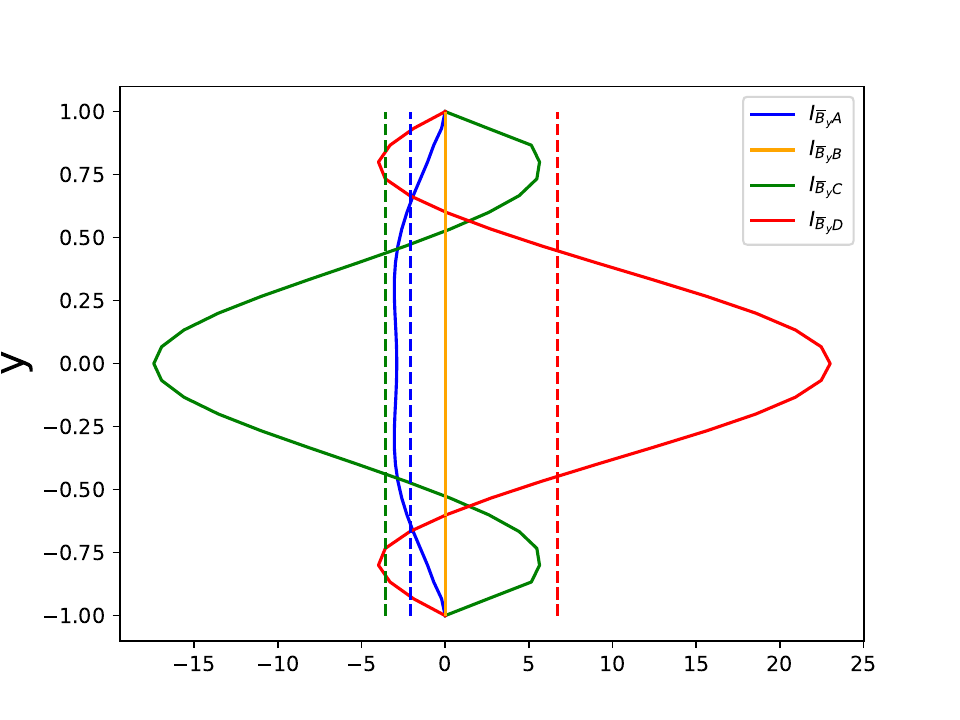}
}
\caption{
Equilibrium balance for the mean poloidal field energy, $\frac{\overline{B}_y^2}{2}$, showing the dominant balance consists of a positive contribution from fluctuation-fluctuation tilting/stretching, $(I_{\overline{B}_yB})$, counteracted by a negative contributions from fluctuation-fluctuation advection, $(I_{\overline{B}_y C})$, and dissipation, $(I_{\overline{B}_y A})$.
Parameters for the fixed-point equilibrium shown are  
$\frac{\nu}{\eta}=0.025$, $\epsilon_{\mybv{u}'\mybv{u}'}=0.4,~ \epsilon_{\mybv{B}'\mybv{B}'}=0,~kx=6.25$.
}
\label{Bybalancefixedpoint}
\end{figure}

Equation for the mean poloidal field, $\overline{B}_y$, is written as:\\ 

\begin{equation}
\begin{split}
\frac{\partial \overline{B}_y}{\partial t}=-\overline{u'_y \frac{\partial B_y'}{\partial y}}-\overline{u_x'\frac{\partial B_y'}{\partial x}}+\overline{B}_y\frac{\partial \overline{u}_y}{\partial y}\\+\overline{B_x'\frac{\partial u'_y}{\partial x}}+ \overline{B_y'\frac{\partial u'_y}{\partial y}}+ \eta \Delta \overline{B}_y-r_B \overline{B}_y
\label{Byeqformaintenance}
\end{split}
\end{equation}\\

The equation for the mean poloidal field energy is obtained by multiplying \eqref{Byeqformaintenance} by $\overline{B}_y$:\\

\begin{equation}
\begin{split}
\frac{\partial \frac{1}{2}(\overline{B}_y)^2}{\partial t}=\overline{B}_y\cdot (-\overline{u'_y \frac{\partial B_y'}{\partial y}}-\overline{u_x'\frac{\partial B_y'}{\partial x}})\\+\overline{B}_y\cdot (\overline{B}_y\frac{\partial \overline{u}_y}{\partial y})+\overline{B}_y\cdot(\overline{B_x'\frac{\partial u'_y}{\partial x}}+ \overline{B_y'\frac{\partial u'_y}{\partial y}})\\+ \overline{B}_y\cdot (\eta \Delta \overline{B}_y-r_B \overline{B}_y)
\end{split}
\end{equation}\\
Contributions to the mean poloidal field energy are:\\

\begin{equation}
I_{\overline{B}_yA}=(\overline{B}_y)\cdot (\eta \Delta \overline{B}_y-r_{B}\overline{B}_y)
\end{equation}
\begin{equation}
I_{\overline{B}_yB}=(\overline{B}_y)\cdot (\overline{B}_y \frac{\partial \overline{u}_y}{\partial y})
\end{equation}
\begin{equation}
I_{\overline{B}_yC}=(\overline{B}_y)\cdot (-\overline{u'_x\frac{\partial B'_y}{\partial x}}-\overline{u'_y\frac{\partial B'_y}{\partial y}})
\end{equation}
\begin{equation}
I_{\overline{B}_yD}=(\overline{B}_y)\cdot (\overline{B'_x\frac{\partial u'_y}{\partial x}}+\overline{B'_y\frac{\partial u'_y}{\partial y}})
\end{equation}\\

The contributions to the mean poloidal field energy forcing are identified as: dissipation, $(I_{\overline{B}_yA})$; mean tilting/stretching, $(I_{\overline{B}_yB})$; fluctuation-fluctuation advection contribution, $(I_{\overline{B}_yC})$; fluctuation-fluctuation tilting/stretching, $(I_{\overline{B}_yD})$.\\

Equilibrium balance maintaining the mean poloidal field energy, $\frac{1}{2}(\overline{B}_y)^2$ for a fixed point equilibrium at $\frac{\nu}{\eta}=0.25$,$\epsilon_{\mybv{u}'\mybv{u}'}=0.4,\epsilon_{\mybv{B}'\mybv{B}'}=0,kx=6.25$ is shown in figure 
\ref{Bybalancefixedpoint}. The dominant balance consists of a positive contribution from fluctuation-fluctuation tilting/stretching, $(I_{\overline{B}_yB})$, counteracted by negative contributions from fluctuation-fluctuation advection, $(I_{\overline{B}_y C})$ and dissipation, $(I_{\overline{B}_y A})$.

\section{Dynamics of Time Dependent ZJTFS}

\begin{figure}

       \centering{%
            \includegraphics[width=.75\linewidth]{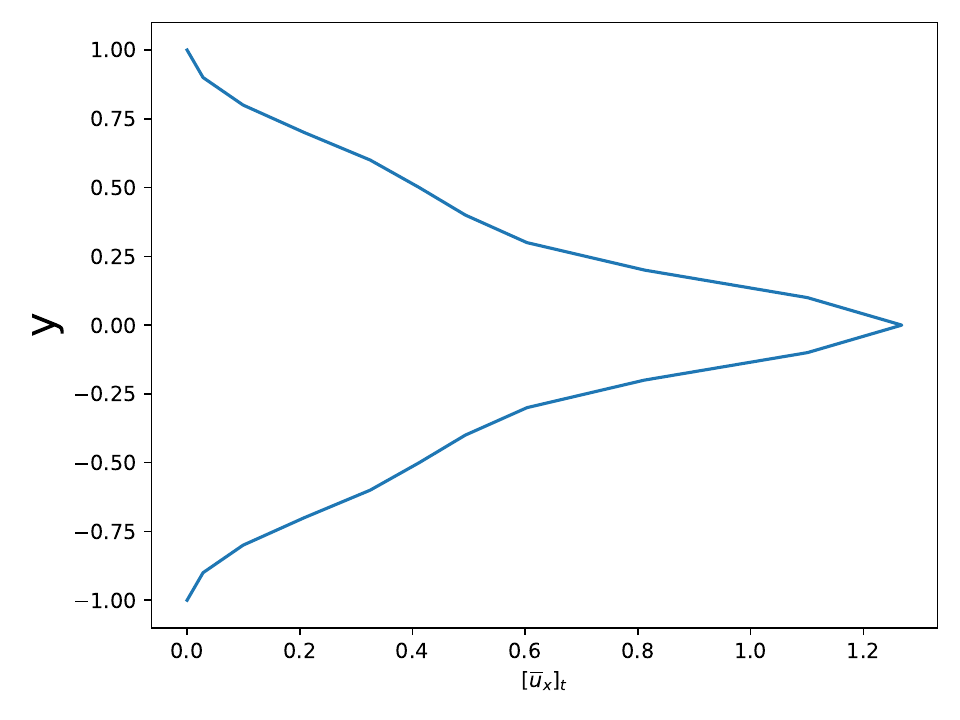}
            
        }

\caption{Structure of the zonal jet equilibrium for parameter values: $\nu=0.0036$, $\epsilon_{\mybv{u}'\mybv{u}'}=0.104$, $\epsilon_{\mybv{B}'\mybv{B}'}=0$
}
\label{fig:solarcycleUequilibria}
\end{figure}

In this section, we employ the S3T SSD framework to investigate the dynamics of the ZJTFS. Our goal is to gain insight into phenomena such as the 22-year solar cycle by analyzing the large-scale dynamo instability of the equilibrium state $(\mybv{\Gamma}_e, \mybv{C}_e)$ for physically relevant parameter values.  Direct observations of the solar cycle velocity and magnetic fields at the photosphere together with  helioseismic inference imply that the cycle is at least reflective of dynamics at the depth of the solar tachocline; indeed, previous studies suggest that the tachocline serves as the fundamental engine of the solar cycle \citep{Tobias2007, Mandal2026, Vorontsov2002}.  We threfore choose parameters based on observational evidence suggesting a characteristic zonal velocity of $1-4m/s$ at tachocline depth \citep{Mandal 2026,Gizon2020, Hathaway2003} choosing as characteristic speed $3m/s$ and as characteristic length 
$L \approx 2 \cdot 10^{8} m \approx 0.3 \cdot  R \odot $.
which implies a characteristic time scale, $\tau \approx 2.2$, years.
Our one layer shallow water model uses reduced gravity instead of full gravity. To represent the dynamics of the solar tachocline, we choose $g=11.1$ and $\beta=0.143$ \citep{Zaqarashvili2009} and
dissipation parameters  $\nu_u=0.0036$, $\nu_h=0$, $r_u=r_h=0.1,r_B=0.2$.
In the preceding section,  we found that choosing $\mybv{F}_{\mybv{u}'}$ of the form \eqref{Fueq}, \eqref{Fpsieq} leads to a finite‑amplitude $\mybv{\overline{B}}$ containing equilibrium that is antisymmetric about 
$y=0$ in the $\overline{B}_x$ component and dipolar in the 
 $\overline{B}_y$ component, resembling that observed in the solar cycle. We wish to preserve these structural features  in  $\overline{B}_x$ and $\overline{B}_y$ while instigating a   time-dependence in the solution to address the solar cycle.  We therefore choose 
$F^{\psi}$ as follows:\\ 
\begin{equation}
F^{\psi}(y_i,y_j)=exp(-3|y_i|)exp(-\frac{(y_i-y_j)^2}{\delta_f^2})
\end{equation}
which excites only divergent free component of $u'_x$ and $u'_y$ in the similar form as $\eqref{Fpsieq}$ while limiting the stochastic excitation primarily near the equator, consistent with the solar cycle. $\mybv{Q}$ has been normalized such that $trace(\mybv{Q})=6.85$. We begin our analysis by finding ZJ equilibria $(\mybv{\Gamma}_e, \mybv{C}_e$) with no magnetic field. 
Shown in figure \ref{fig:solarcycleUequilibria} is the structure of the mean zonal velocity component for the ZJ equilibrium  at $\nu=0.0036, ~\epsilon_{\mybv{u}'\mybv{u}'}=0.09,~ \epsilon_{\mybv{B}'\mybv{B}'}=0$. The structures shown in the figure are consistent with those reported in earlier SSD based studies of sharp jets in a low-dissipation regime
\citep{Farrell 2003}. We note that fixed-point ZJ equilibria  support a manifold of Rossby–gravity waves.

We now examine the large scale dynamo instability of the fixed-point jet equilibria, $(\mybv{\Gamma}_e, \mybv{C}_e)$, shown in figure \ref{fig:solarcycleUequilibria} using equations \ref{S3Tstabilitymeaneq} and \ref{S3Tstabilitycoveq}. 

\begin{figure}
\centering{
        \subfloat[]{%
            \includegraphics[width=.75\linewidth]{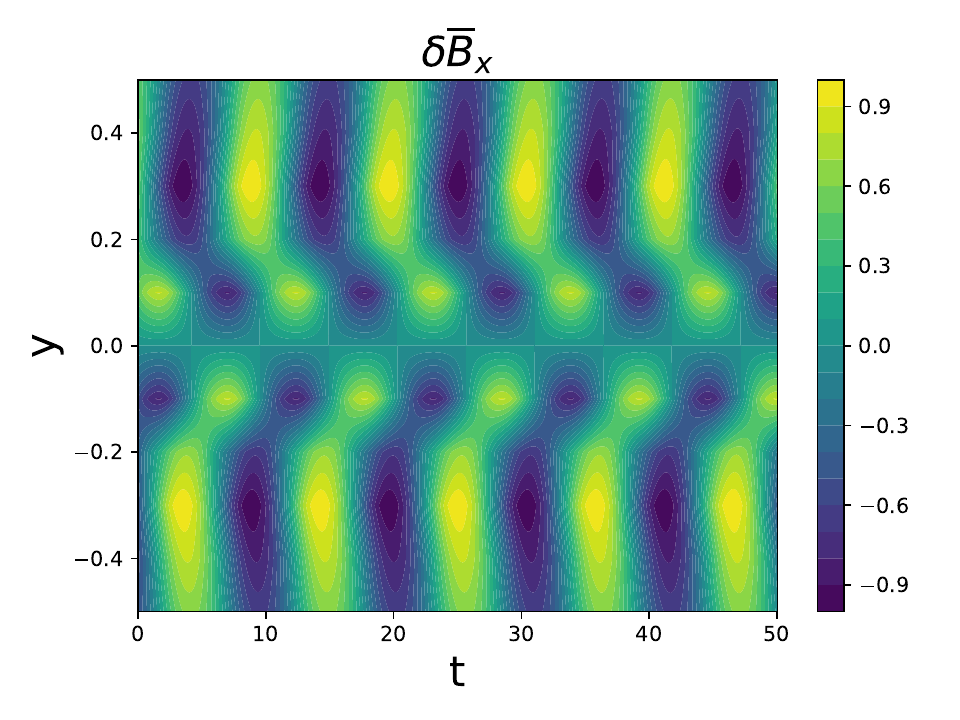}
            \label{subfig:1b}%
        }\\
        \subfloat[]{%
            \includegraphics[width=.75\linewidth]{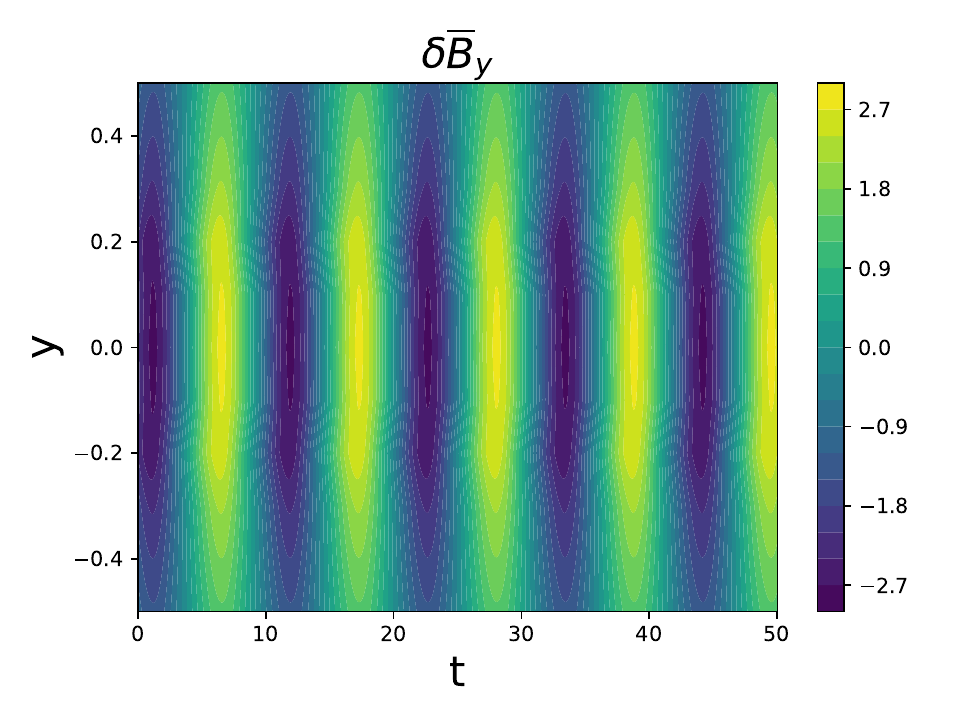}
            \label{subfig:1c}%
        }
        }

\caption{Structure of large scale dynamo instabilty shown as a Hovmöller diagram. panel $(a)$: mean toroidal magnetic field $\delta \overline{B}_x$. Panel $(b)$: mean poloidal magnetic field $\delta \overline{B}_y$. This eigenmode has been normalized such that the maximum value of $\delta \overline{B}_x$ is unity. $\nu=0.0036$, $\eta= 0.36$, $\frac{\nu}{\eta}=0.01$, $\epsilon_{\mybv{u}'\mybv{u}'}=0.09$, $\epsilon_{\mybv{B}'\mybv{B}'}=0$. growth rate $Re(\sigma)=0.23$, frequency $Im(\sigma)=0.5712$}
\label{FloquetS3Tmode}
\end{figure}
\begin{figure}

       \centering {
            \includegraphics[width=.75\linewidth]{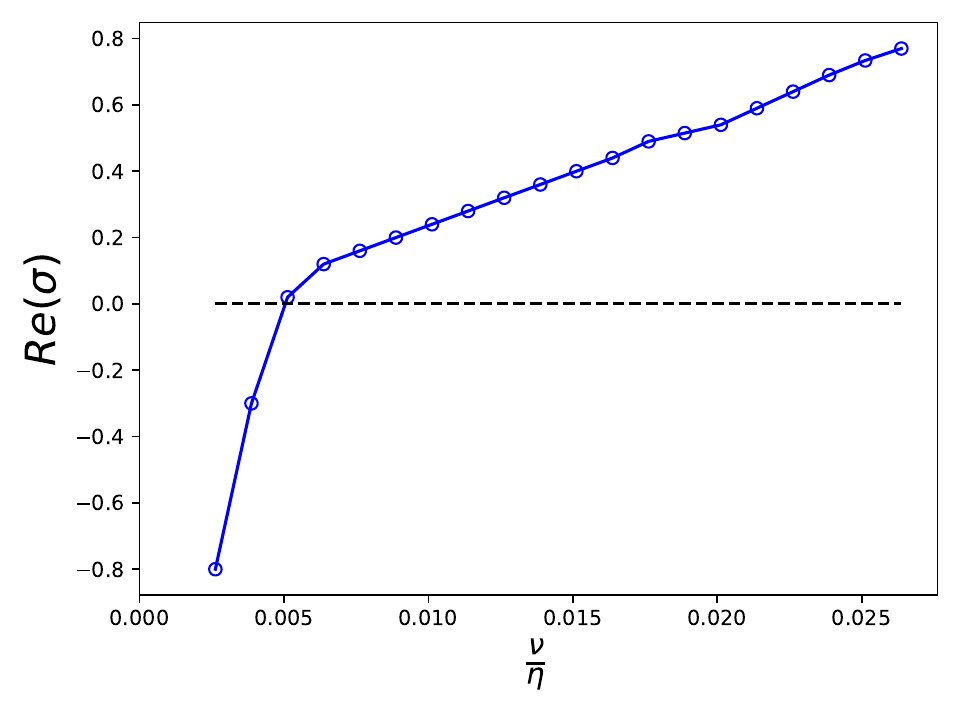}}

\caption{Stability diagram for large scale dynamo instability of the zonal jet equilibria showing growth rate $Re(\sigma)$  as function of Prandtl number $Pr=\frac{\nu}{\eta}$. Black dashed line indicates neutral stability
}
\label{Prandtlvacillatinginstability}
\end{figure}
\begin{figure}

       \centering {
            \includegraphics[width=.75\linewidth]{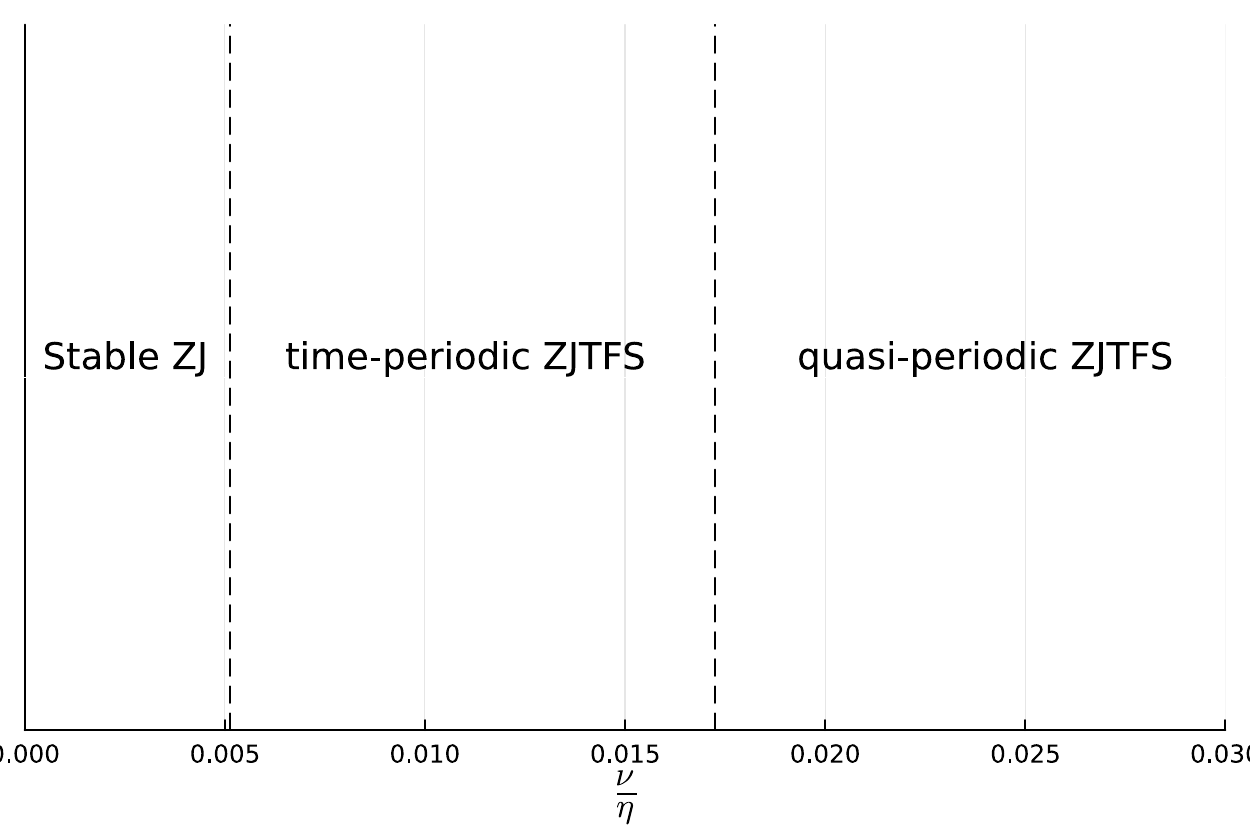}
     
      }
\caption{Equilibrium regime diagram as a function of $\frac{\nu}{\eta}$ for $\nu=0.0036$ and $\epsilon_{\mybv{u'}\mybv{u}'}=0.09,\epsilon_{\mybv{B}'\mybv{B}'}=0$. $R_0$: stable regime( $\overline{B}_x(y,t)=0$).$R_1$: time periodic equilibria with equatorial antisymmetry in $\overline{B}_x$. $R_2:$ time periodic equilibria with broken equatorial antisymmetry in $\overline{B}_x$. $R_3$: Limit Cycle)
}
\label{Prandtlregimevacillation}
\end{figure}
\begin{figure}
\centering{
 \subfloat[]{
    
            \includegraphics[width=.75\linewidth]{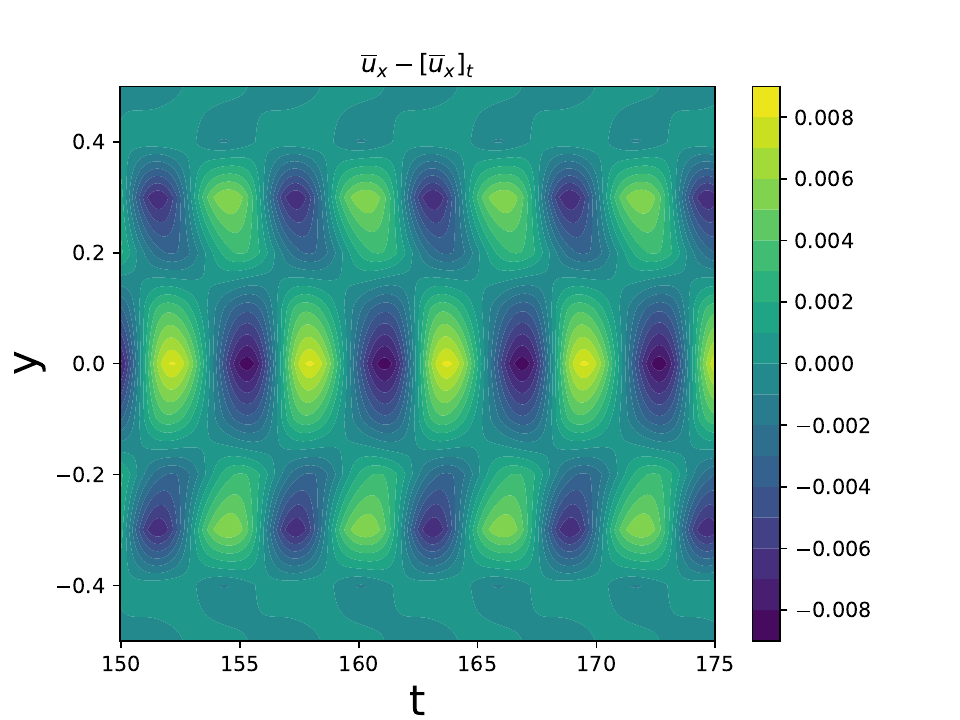}
           \label{subfig:1b}%
        
        }\\
         \subfloat[]{

         \includegraphics[width=.75\linewidth]{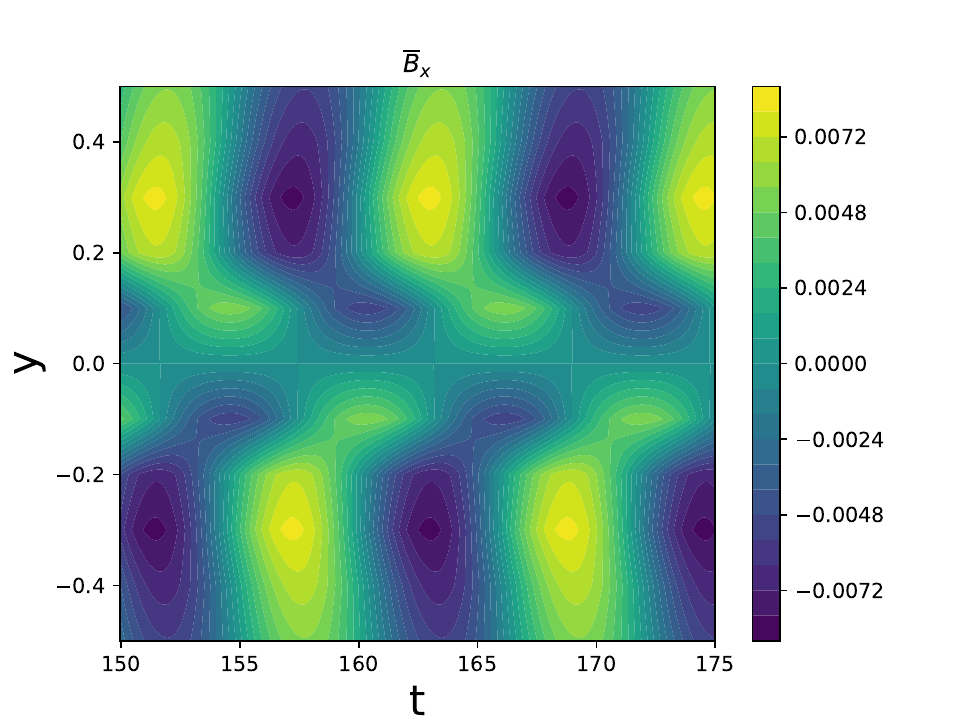}
         \label{subfig:1c}%
         }
         }

\caption{Structure of a time-periodic ZJTFS shown as a Hovmöller diagram. Panel (a): $\overline{u}_x(y,t)$ with equatorial symmetry. Panel (b): $\overline{B}_x(y,t)$ with equatorial antisymmetry.
 $\frac{\nu}{\eta}=0.006;\nu=0.0036$, $\epsilon_{\mybv{u}'\mybv{u}'}=0.09$, $\epsilon_{\mybv{B}'\mybv{B}'}=0$ }
\label{R1hovmoller}
\end{figure}
\begin{figure}

\centering{
 \subfloat[]{
    
            \includegraphics[width=.75\linewidth]{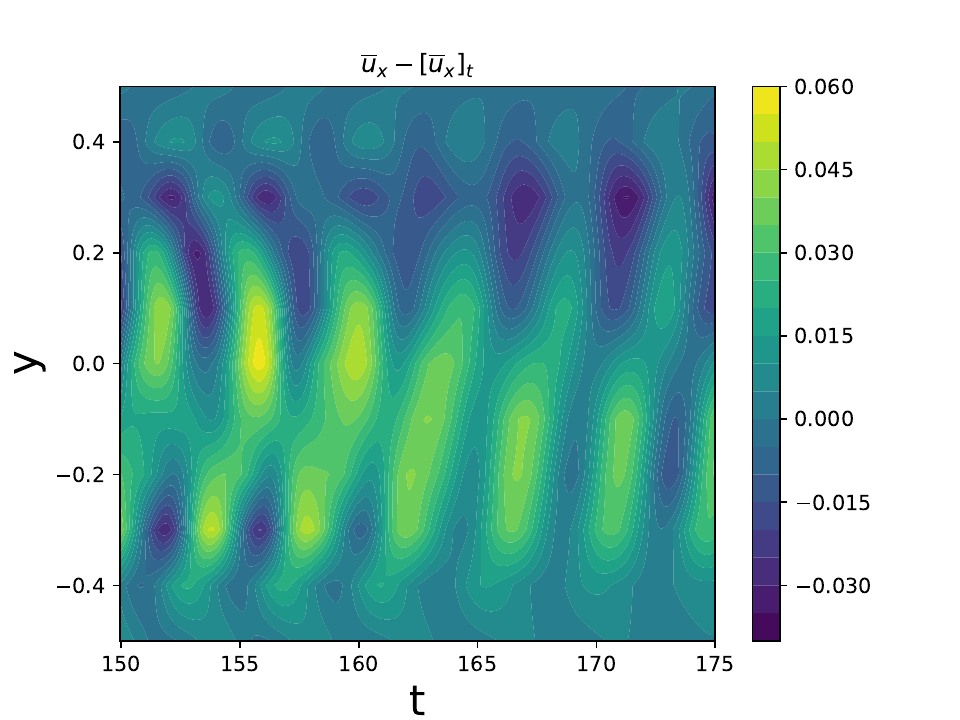}
           \label{subfig:1b}%
        
        }\\
         \subfloat[]{

         \includegraphics[width=.75\linewidth]{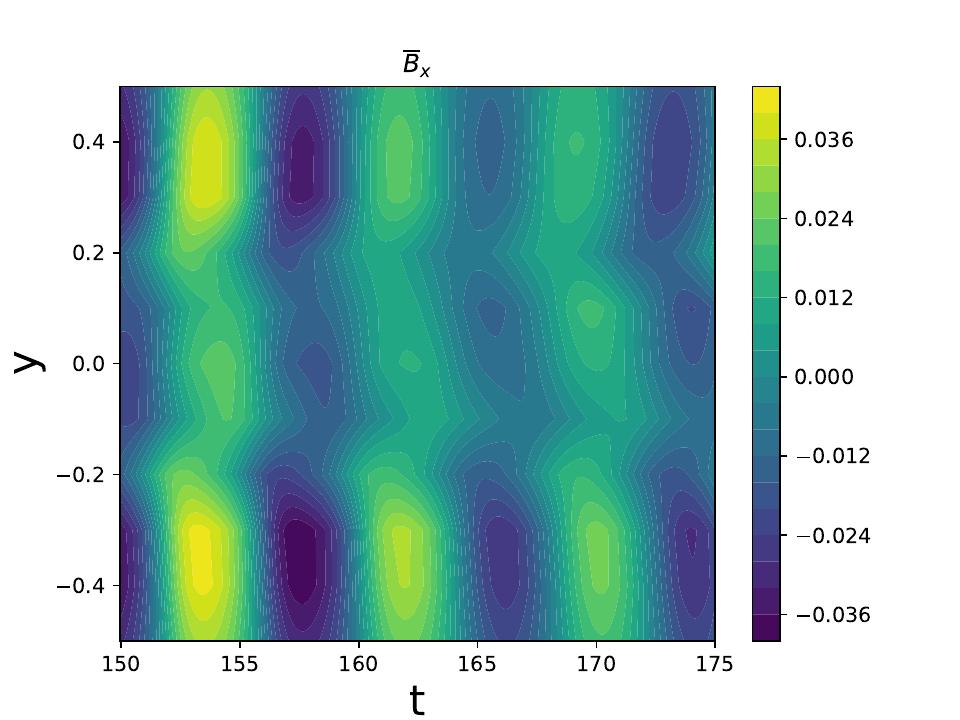}
         \label{subfig:1c}%
         }
         }

\caption{Structure of a quasi-periodic ZJTFS equilibrium shown as a Hovmöller diagram. Panel (a): $\overline{u}_x(y,t)$ with broken equatorial symmetry. Panel (b): $\overline{B}_x(y,t)$ with broken equatorial antisymmetry. $\frac{\nu}{\eta}=0.018;\nu=0.0036$, $\epsilon_{\mybv{u}'\mybv{u}'}=0.09$, $\epsilon_{\mybv{B}'\mybv{B}'}=0$}
\label{R2hovmoller}
\end{figure}

We vary the magnetic Prandtl number,  $\frac{\nu}{\eta}$, as a stability parameter.
Shown in figure $\ref{FloquetS3Tmode}$ are Hovmöller diagrams of the mean magnetic field components, $\delta \overline{\mybv{B}}$, of the unstable eigenmode  for Prandtl number  $\frac{\nu}{\eta}=0.01$. This eigenmode exhibits spatiotemporal structure simliar to that of the 22 year old solar cycle, known for its characteristic “butterfly” shape. Considering that characteristic time scale is around $2.2$ years in our S3T SSD formulated dynamics, the periodicity of approximately 11 dimensionless time units, or frequency $\omega=0.57$, shown in the figure corresponds to a physical period of about 24.2 years. Exact periodicity of the unstable eigenmode depends on parameters including $\epsilon_{\mybv{u}'\mybv{u}'}$, $\nu_u$ , $r_u, r_B$. 
Shown in figure \ref{Prandtlvacillatinginstability} is a stability diagram for the large scale dynamo mode supported by the ZJ equilibrium shown in figure \ref{fig:solarcycleUequilibria}, as a function of Prandtl number $\frac{\nu}{\eta}$.  The ZJ becomes marginally stable to large scale dynamo instability at $\frac{\nu}{\eta}\approx 0.0051$.
The S3T SSD is a nonlinear system with an equilibration mechanism consistent with its underlying MHDSW dynamics. Perturbing the zonal jet equilibrium with an unstable S3T mode results in establishing a finite-amplitude ZJTFS equilibrium.  This equilibrium may consist of a fixed-point, a time-periodic, or a quasi-periodic statistical equilibrium. As a summary of our findings and to set the stage for the presentation of our results, an equilibrium structure diagram as a function of
the Prandtl number $Pr=\frac{\nu}{\eta}$ is shown in figure
\ref{Prandtlregimevacillation}.  This figure shows that the fixed-point ZJ is stable to large scale dynamo formation, as indicated by  no $\overline{B}_x$ component arising, for $\frac{\nu}{\eta}< 0.0051$. Around $\frac{\nu}{\eta} \approx 0.0051$, ZJ equilibria becomes unstable to large scale dynamo formation. For moderately unstable eigenmode at $0.0051<\frac{\nu}{\eta} < 0.017$, it equilibrates as a time-periodic ZJTFS equilibrium. Time-periodic equilibria exhibit equatorial antisymmetry in $\overline{B}_x$ and equatorial symmetry in $\overline{u}_x$. A representative Hovmöller diagram of a time-periodic ZJTFS at $\frac{\nu}{\eta}=0.006$ is shown in figure \ref{R1hovmoller}. 
When $\frac{\nu}{\eta}$ increases further so $ \frac{\nu}{\eta}>0.017$, the ZJ equilibrates to a quasi-periodic ZJTFS equilibrium. Quasi-periodic equilibria have broken equatorial symmetry in $\overline{u}_x$ and broken equatorial antisymmetry in $\overline{B}_x$. 
A representative Hovmöller diagram of a quasi-periodic ZJTFS equilibria at $\frac{\nu}{\eta}=0.018$ is shown in figure \ref{R2hovmoller}. 

\section{Mechanisms Maintaining the time-periodic ZJTFS}
The dynamics of formation and maintenance of time-dependent ZJTFS  has been captured by the S3T SSD implementation of SWMHD dynamics. 
The time-periodic ZJTFS equilibria regime, which contains equatorial anti-symmetry in $\overline{B}_x(y,t)$, is most relevant to the solar cycle.  In this section, we analyze the mechanisms maintaining the mean velocity and magnetic field components in this regime in order to gain insight into solar cycle dynamics. 
 
\begin{figure}
\centering{
\includegraphics[width=\linewidth,height=0.375\textheight]{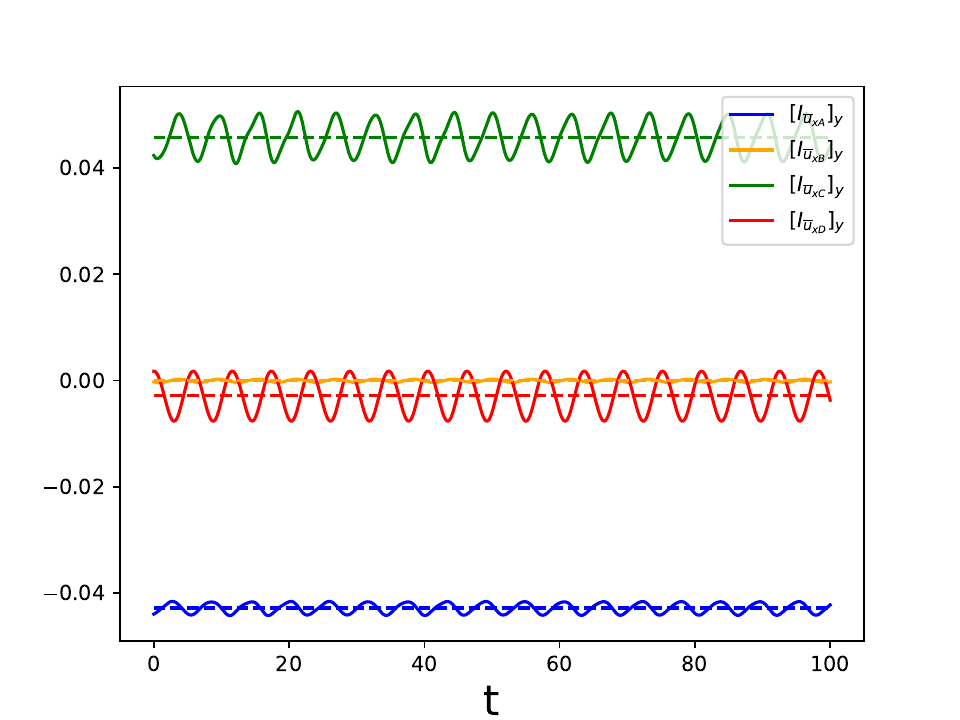}
}
\caption{
Equilibrium balance for mean zonal kinetic energy, $\frac{\overline{u}_x^2}{2}$, showing dissipation, $(I_{\overline{u}_xA})$, mean magnetic tension force contribution, $(I_{\overline{u}_xB})$, Reynolds stress force contribution, $(I_{\overline{u}_xC})$, and the fluctuation-fluctuation magnetic tension force contribution, $(I_{\overline{u}_xD})$. Terms are averaged in y. The time-periodic ZJTFS has parameters  $\frac{\nu}{\eta}=0.006$, $\epsilon_{\mybv{u}'\mybv{u}'}=0.09, \epsilon_{\mybv{B}'\mybv{B}'}=0$. 
}
\label{uxmaintenancePr0076}
\end{figure}

The time series of the components maintaining the mean zonal kinetic energy, $\frac{1}{2} (\overline{u}_x)^2$, for a time-periodic ZJTFS equilibrium at $\frac{\nu}{\eta}=0.006$ are shown in figure 
\ref{uxmaintenancePr0076}. 
The dominant balance consists of a positive contribution from  Reynolds stress, $(I_{\overline{u}_xC})$, opposed by a negative contribution from dissipation, $(I_{\overline{u}_xA})$, augmented by minor contributions from fluctuation-fluctuation magnetic tension,  $(I_{\overline{u}_xD})$. 

 \begin{figure}
\centering{
\includegraphics[width=\linewidth,height=0.375\textheight]{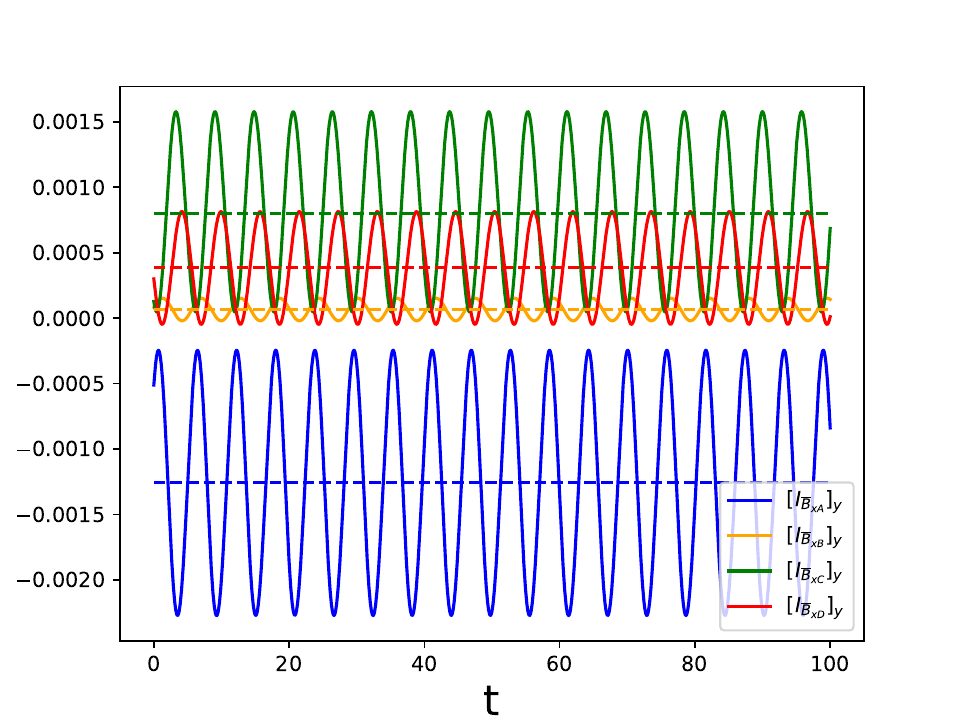}
}
\caption{
Equilibrium balance for the mean toroidal field energy, $\frac{\overline{B}_x^2}{2}$ , showing the dissipation, $(I_{\overline{B}_xA})$, the mean tilting/stretching, $(I_{\overline{B}_xB})$, the fluctuation-flucutation advection contribution, $(I_{\overline{B}_xC})$, and the fluctuation-fluctuation tilting/stretching, $(I_{\overline{B}_xD})$. Terms are averaged in y.
A time-periodic ZJTFS is at  $\frac{\nu}{\eta}=0.006$, $\epsilon_{\mybv{u}'\mybv{u}'}=0.09, \epsilon_{\mybv{B}'\mybv{B}'}=0$.
}
\label{BxmaintenancePr0076}
\end{figure}

 \begin{figure}
\centering{
\includegraphics[width=\linewidth,height=0.375\textheight]{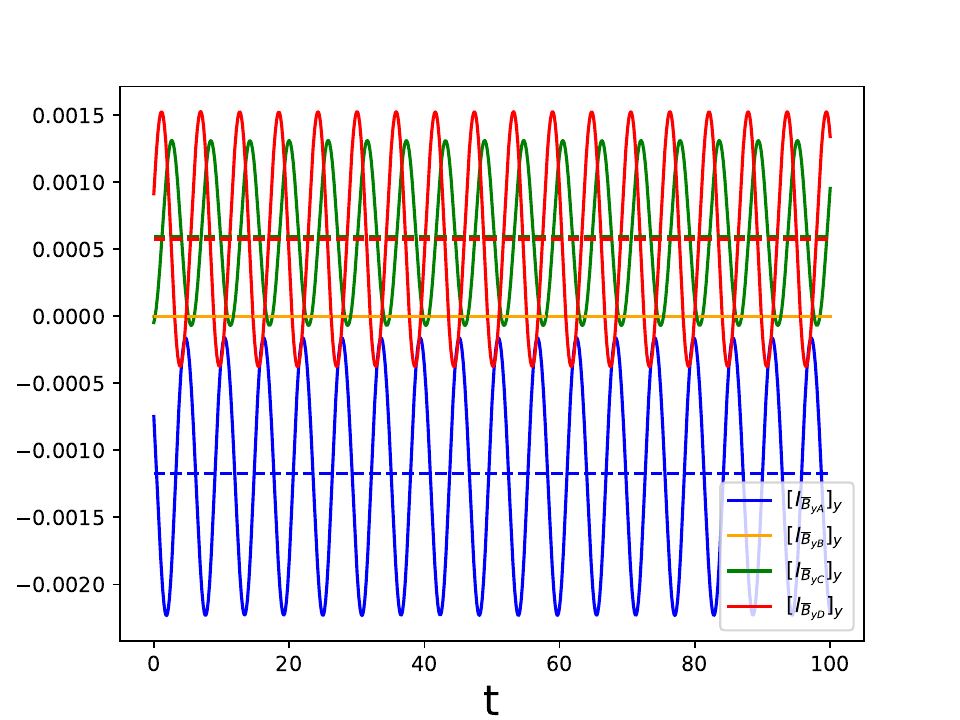}
}
\caption{
Equilibrium balance for the mean poloidal field energy, $\frac{\overline{B}_y^2}{2}$ , showing the dissipation, $(I_{\overline{B}_yA})$, the mean tilting/stretching, $(I_{\overline{B}_yB})$, the fluctuation-fluctuation advection contribution, $(I_{\overline{B}_yC})$, and the fluctuation-fluctuation tilting/stretching, $(I_{\overline{B}_yD})$. Terms are averaged in y.
A time-periodic ZJTFS is at  $\frac{\nu}{\eta}=0.006$, $\epsilon_{\mybv{u}'\mybv{u}'}=0.09,~ \epsilon_{\mybv{B}'\mybv{B}'}=0$.
}
\label{BymaintenancePr0076}
\end{figure}
The time series of components maintaining the mean toroidal field energy, $\frac{1}{2} (\overline{B}_x)^2$, for the same case is shown in Figure \ref{BxmaintenancePr0076}. 
The dominant balance is between fluctuation-fluctuation advection, $(I_{\overline{B}_xC})$, and fluctuation-fluctuation tilting/stretching, $(I_{\overline{B}_xD})$ with a small additional positive contribution coming from mean tilting/stretching, $(I_{\overline{B}_xB})$.  These positive contributions are balanced by dissipation, $(I_{\overline{B}_xA})$.
The positive contributions of $(I_{\overline{B}_xC})$ and $(I_{\overline{B}_xD})$ to torroidal field forcing imply upgradient energy transfer from $\mybv{u}'$ and $\mybv{B}'$ to $\overline{B}_x$.\\

The time series of the components maintaining the mean poloidal field energy, $\frac{1}{2} (\overline{B}_y)^2$, is shown in figure 
\ref{BymaintenancePr0076}. 
The dominant balance consists of  positive contributions from fluctuation-fluctuation advection, $(I_{\overline{B}_yC})$, and fluctuation-fluctuation tilting/stretching, $(I_{\overline{B}_yD})$, balanced by a negative contribution from dissipation, $(I_{\overline{B}_yA})$. 
The positive contributions of $(I_{\overline{B}_yC})$ and $(I_{\overline{B}_yD})$ imply a  higher energy transfer from $\mybv{u}'$ and $\mybv{B}'$ to $\overline{B}_y$.\\
Our maintenance diagnostics of $\overline{B}_x$ and $\overline{B}_y$ reveal that tilting in both directions (poloidal to toroidal, and toroidal and poloidal) contributes to mean magnetic field maintenance. However, tilting/stretching occurs predominantly through the fluctuation–fluctuation tilting/stretching terms ($I_{\overline{B}_xD}$) and  ($I_{\overline{B}_yD}$). This explicit identification of the mechanism providing the dominant balance that maintains the poloidal field differs from the traditional parameterization of this turbulent transfer by a factor $\alpha$ in the induction equation.

\section{Discussion and Conclusion}

Previous studies have addressed the ubiquity of zonal jet structures in planetary and stellar turbulence by applying SSD methods, including studies that used the shallow water equatorial $\beta$ plane S3T SSD to explain the banded winds of the gas giants \citep{Farrell 2009}.  The observed coexistence of velocity jets and coherent large-scale magnetic fields in stellar turbulence motivates extending this hydrodynamic analysis method to the magnetohydrodynamic regime.
For the solar cycle, while mean toroidal and poloidal magnetic fields coexist with a mean zonal velocity, neither a specific instability forming the zonal velocity field directly from a turbulent state nor one forming the mean $\overline{\mybv{B}}$ field directly from a turbulent state has been previously identified. The reason  is that the instabilities forming these structures are nonlinear. 
The key to discovering these  instabilities lies in analyzing the MHD turbulence dynamics using an SSD formulation of MHD. In the S3T SSD for MHD closed at second order the second cumulant, which is the ensemble mean covariance of the state fluctuations, is treated as a linear variable in the perturbation analysis, although it is quadratically nonlinear as a state variable, which allows the full power of linear perturbation analysis to be brought to bear on the associated nonlinear instability.\\

Previously, the perturbative form of S3T SSD has been employed to demonstrate that, given a background turbulent field, a ZJ emerges as an instability with bifurcation parameter the turbulence excitation level \citep{Farrell 2003, Farrell 2009}.  A finite amplitude fixed-point or time-dependent (vacillating) jet equilibrium proceeds from this instability.  In this work, we employed perturbation form of S3T SSD to investigate large scale dynamo instability supported by this finite amplitude ZJ equilibria. We identify the Prandtl number, $Pr=\frac{\nu}{\eta}$, as the bifurcation parameter above which this finite amplitude ZJ becomes unstable to large scale dynamo instability  which in turn equilibrates to a finite amplitude ZJTFS.


For ZJ equilibria at parameter values relevant to the solar cycle, the spatiotemporal structure of the mean toroidal field—both during the large-scale dynamo instability and within the subsequent time-periodic ZJTFS equilibria—closely resembles that of the solar cycle.

By employing the nonlinear S3T SSD framework, we can analyze both the dynamics underlying the onset of large-scale dynamo formation as an instability and that of the nonlinear ZJTFS that arise from these unstable modes. The regime most relevant to the 22-year solar cycle is the time-periodic ZJTFS equilibrium, which exhibits equatorial symmetry and antisymmetry in $\overline{u}_x$ and $\overline{B}_x$, respectively. Consequently, we focus on the maintenance mechanisms within this regime.  We find that the velocity jet ($\overline{u}_x$) is primarily maintained by Reynolds stress contributions, which are balanced by viscous dissipation with a minor negative contribution from magnetic tension.  
Turning our attention to the $\overline{B}_x$ component, we find that it is maintained dominantly by fluctuation-fluctuation  advection and fluctuation-fluctuation tilting balanced by dissipation. Lastly, $\overline{B}_y$ component is maintained dominantly by fluctuation-fluctuation terms balanced by dissipation.  

In comparison with classical dynamo theory we replace the traditional $\alpha$-$\omega$ dynamo instability model, which
relies on a parameterized scalar, $\alpha$, to regenerate a poloidal field from the toroidal field, with the SSD instability mechanism which explicitly resolves the feedback between the mean state and the fluctuation covariance. 
In this SSD instability, the mode associated with the amplification of the toroidal field by $\omega$-tilting of the poloidal field relies on fluctuation-fluctuation components to produce the poloidal field closing the dynamo loop.  

However, we find that the $\alpha - \omega$ dynamo mechanism plays a minor role in maintaining the toroidal field at parameter regimes relevant for solar cycle dynamics.  The dominant contributions to maintaining toroidal field in the solar cycle is rather direct forcing by the finite amplitude fluctuation-fluctuation terms.

In summary, in one unified framework, starting from a turbulent, rotating, and conducting fluid, the S3T SSD allows analytical identification of a nonlinear instability that produces a ZJ of finite-amplitude. This ZJ, in turn, supports a secondary nonlinear dynamo instability resulting in a finite-amplitude toroidal and poloidal field that coexists with the ZJ. With an appropriate selection of parameters, this ZJTFS recovers the spatiotemporal structure and time dependence of the solar cycle in both its velocity and magnetic fields.

\appendix
\label{sec:appendix}
\section{The Individual Components of $\mybv{A}$}
The matrix of the dynamics, $\mybv{A}$, is defined as:
\label{sec:Aindividual}
\begin{equation} 
\mybv A= \begin{bmatrix}A_{11} && A_{12} && A_{13} && A_{14} && A_{15} \\ A_{21} && A_{22} &&A_{23} && A_{24} && A_{25} \\ A_{31} && A_{32} && A_{33} && A_{34} && A_{35} \\ A_{41}  && A_{42} && A_{43} && A_{44} && A_{45} \\ A_{51} && A_{52} && A_{53} && A_{54} && A_{55}\end{bmatrix}
\end{equation}\\
The individual components are defined as:

\begin{equation} \begin{split} \\ A_{11}=- \overline{u}_x(ik_x)-\overline{u}_y\partial_y+ \nu_{u} (\partial_{yy}-k_x^2)-r_{u} \\
A_{12}=-\frac{\partial \overline{u}_x}{\partial y}+f \\ A_{13}= -g (ik_x) \\ A_{14}= \overline{B_x}(ik_x) + \overline{B_y} \partial_y \\ A_{15}=  \frac{\partial \overline{B_x}}{\partial y} \\ A_{21}= -f \\ A_{22}= -\overline{u}_x(ik_x)-\overline{u}_y\partial_y  -\frac{\partial \overline{u}_y}{\partial y}+ \nu_{u} (\partial_{yy}-k_x^2)-r_u \\ A_{23}= -g \partial_y \\ A_{24}= 0 \\ A_{25}= \overline{B}_x (ik_x)+ \overline{B}_y\partial_y+ \frac{\partial \overline{B}_y}{\partial y} \\ A_{31}= -(ik_x) \overline{h} \\ A_{32}= -\frac{\partial \overline{h}}{\partial y}-\overline{h}\partial_y \\ A_{33}= -(ik_x)\overline{u} -\overline{u}_y \partial_y -\frac{\partial \overline{u}_y}{\partial y} + \nu_h (\partial_{yy}-k_x^2)-r_h \\ A_{34}=0 \\ A_{35}=0  
 \\ A_{41}=
    (\overline{B}_x(ik_x)+\overline{B_y}\partial_y) \\ A_{42}=  -\frac{\partial \overline{B_x}}{\partial y} \\ A_{43}=0  \\ A_{44}= -\overline{u}_x (ik_x)-\overline{u}_y\partial_y-r_B \\ A_{45}= \frac{\partial \overline{u}_x}{\partial y} \\ A_{51}=0 \\A_{52}= (\overline{B_x}(ik_x)+\overline{B_y}\partial_y)-\frac{\partial \overline{B_y}}{\partial y} \\ A_{53}=0 \\ A_{54}=0 \\ A_{55}= -\overline{u}_x (ik_x)-\overline{u}_y\partial_y+ \frac{\partial \overline{u}_y}{\partial y}-r_B
\end{split}
\end{equation}

\section{The Invidiual Components of $\mybv{G}(\mybv{\Gamma})$ and $\mybv{L}(\mybv{C})$}
\label{sec:Ggammaindividual}
The individual components of $\mybv{G}(\mybv{\Gamma})$ and  $\mybv{L}(\mybv{C})$ are defined as:
\begin{equation}
\mybv{G}(\mybv \Gamma)=\begin{bmatrix}-\overline{u}_y\frac{\partial \overline{u}_x}{\partial y}+\frac{\partial \overline{B}_x}{\partial y}\overline{B}_y+ \nu_{u} \Delta \overline{u}_x-r_u \overline{u}_x+ (f_0+ \beta y) \overline{u}_y \\ -\overline{u}_y\frac{\partial \overline{u}_y}{
\partial y}-g\frac{\partial \overline{h}}{\partial y}+\frac{\partial \overline{B}_y}{\partial y} \overline{B}_y+ \nu_{u} \Delta \overline{u}_y-r_u \overline{u}_y-(f_0+\beta y) \overline{u}_x\\ -\overline{h}\frac{\partial (\overline{u}_y)}{\partial y}-\overline{u}_y\frac{\partial (\overline{h})}{\partial y}-r_h(\overline{h}-1)+\nu_{h}\Delta \overline{h} \\ -\overline{u}_y\frac{\partial \overline{B_x}}{\partial y}+\overline{B_y}\frac{\partial \overline{u}_x}{\partial y}+ \eta \Delta \overline{B}_x-r_B \overline{B}_x \\ -\overline{u}_y\frac{\partial \overline{B}_y}{\partial y}+\overline{B}_y\frac{\partial \overline{u}_y}{\partial y}+ \eta \Delta \overline{B}_y-r_B \overline{B}_y \end{bmatrix}
\end{equation}

\begin{equation}
\mybv L(\mybv C)= \begin{bmatrix}-\overline{u'_y \frac{\partial u'_x}{\partial y}}+\overline{\frac{\partial B_x'}{\partial y} B_y'} \\ -\overline{u_x'(ik_x) u'_y}-\overline{u'_y\frac{\partial u'_y}{\partial y}}+\overline{(ik_x) B_y'B_x'}+\overline{\frac{\partial B_y'}{\partial y}B_y'} \\ -\frac{\partial \overline{h'u'_y}}{\partial y} \\ -\overline{u'_x (ik_x) B_x'}-\overline{u'_y \frac{\partial B_x'}{\partial y}}+ \overline{B_x'(ik_x) u_x'}+\overline{B_y'\frac{\partial u'_x}{\partial y}} \\ -\overline{u'_x (ik_x) B_y'}-\overline{u'_y \frac{\partial B_y'}{\partial y}}+ \overline{B_x'(ik_x) u'_y}+\overline{B_y'\frac{\partial u'_y}{\partial y}}
 \end{bmatrix}
\end{equation}\\

\end{document}